\documentclass[aps,twocolumn,showpacs,floatfix,superscriptaddress,letterpaper,longbibliography]{revtex4-1}
\usepackage{graphicx}
\usepackage{bm}
\usepackage{amsmath,amsfonts}
\usepackage{amsbsy}
\usepackage[colorlinks=true,linkcolor=blue,urlcolor=blue,citecolor=blue,breaklinks=true]{hyperref}
\usepackage[utf8]{inputenc}
\DeclareUnicodeCharacter{03BC}{\mbox{$\mu$}}
\DeclareUnicodeCharacter{2009}{ }

\begin{document}

\title{Thermal conductivity of the degenerate one-dimensional Fermi gas}

\author{K. A. Matveev}

\affiliation{Materials Science Division, Argonne National Laboratory,
  Argonne, Illinois 60439, USA}

\author{Zoran Ristivojevic}

\affiliation{Laboratoire de Physique Th\'{e}orique, Universit\'{e} de
  Toulouse, CNRS, UPS, 31062 Toulouse, France}

\date{May 6, 2019}

\begin{abstract}
  We study heat transport in a gas of one-dimensional fermions in the
  presence of a small temperature gradient.  At temperatures well
  below the Fermi energy there are two types of relaxation processes
  in this system, with dramatically different relaxation rates.  As a
  result, in addition to the usual thermal conductivity, one can
  introduce the thermal conductivity of the gas of elementary
  excitations, which quantifies the dissipation in the system in the
  broad range of frequencies between the two relaxation rates.  We
  develop a microscopic theory of these transport coefficients in the
  limit of weak interactions between the fermions.
\end{abstract}
\maketitle

\section{Introduction}
\label{sec:introduction}

Relaxation of one-dimensional systems toward equilibrium has a number
of special features.  The two-particle scattering processes, which
control relaxation in higher dimensions, are strongly restricted in
one dimension by the conservation laws, and do not lead to effective
relaxation of the system.  As a result, the relaxation is dominated by
three-particle processes.  In a quantum system at low temperature $T$
these scattering processes are strongly suppressed, resulting in a
slow relaxation toward equilibrium
\cite{imambekov_one-dimensional_2012}.  This leads to a different
temperature dependence of the transport coefficients at low
temperatures.  For example, while the bulk viscosity of the
three-dimensional Fermi liquid vanishes at $T\to0$ as $\zeta\propto
T^2$ \cite{sykes_transport_1970}, in one dimension it grows as
$\zeta\propto T^{-3}$ \cite{matveev_viscous_2017}.

\begin{figure*}
\includegraphics[width=.95\textwidth]{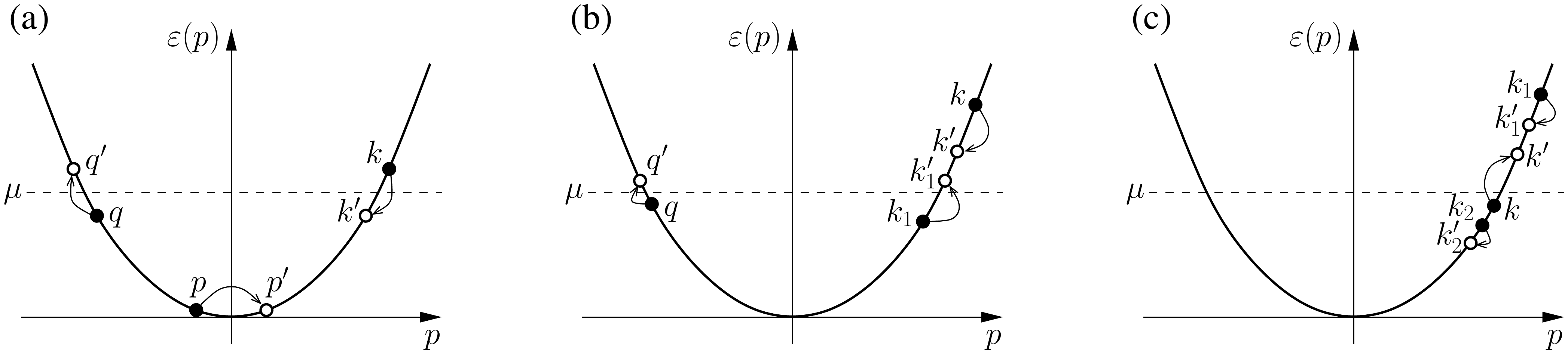}
\caption{The dominant three-particle scattering processes.  Solid
  lines show energy of the particle as a function of momentum, while
  dashed lines show the position of the chemical potential.  Processes
  of type (a) enable the backscattering of fermions; the resulting
  relaxation rate $\tau^{-1}$ is exponentially small.  Processes of
  types (b) and (c) do not change the numbers of the right- and
  left-moving particles. Their rates scale as a power of temperature.}
\label{fig:basic-processes}
\end{figure*}

Another important feature of one-dimensional systems is that each
particle moves in one of only two directions.  As a result, at low
temperatures the dominant scattering processes with small momentum
transfer are very inefficient at changing the direction of motion.
This effect is best illustrated in the case of a one-dimensional Fermi
gas.  The most efficient three-particle process that changes the
relative number of the right- and left-moving particles is shown in
Fig.~\ref{fig:basic-processes}(a).  In order for an electron to change
the direction of motion, this process must involve a hole near the
bottom of the band.  Thus the rate of such processes is exponentially
small, $\tau^{-1}\propto\exp(-\mu/T)$ \cite{lunde_three-particle_2007,
  micklitz_transport_2010, matveev_equilibration_2012,
  matveev_scattering_2012}, where $\mu$ is the chemical potential.  On
the other hand, the scattering processes shown in
Fig.~\ref{fig:basic-processes}(b) and (c) do not change the numbers of
the right- and left-moving particles, but rearrange excitations near
the two Fermi points.  The corresponding rate of scattering of thermal
excitations $\tau_{\rm ex}^{-1}$ scales as a power of temperature.  In
spinless one-dimensional systems $\tau_{\rm ex}^{-1}\propto T^7$
\cite{imambekov_one-dimensional_2012, arzamasovs_kinetics_2014,
  protopopov_relaxation_2014}, while for weakly interacting
spin-$\frac12$ fermions $\tau_{\rm ex}^{-1}\propto T$
\cite{karzig_energy_2010}.

The presence of exponentially slow relaxation processes in the system
results in a very large thermal conductivity $\kappa$ in one
dimension.  Phenomenological treatment
\cite{degottardi_electrical_2015} of the transport in a spinless
one-dimensional quantum liquid based on the Luttinger liquid theory
\cite{haldane_luttinger_1981} yields
\begin{equation}
  \label{eq:kappa_Luttinger}
  \kappa=\frac{\pi Tv\tau}{3\hbar}.
\end{equation}
Here $v$ is the velocity of the bosonic excitations in the Luttinger
model and $\hbar$ is the Planck's constant.  It is important to note
that the above result applies to thermal conductivity measured at low
frequencies $\omega\ll \tau^{-1}$.  At higher frequencies the
exponentially slow relaxation processes of
Fig.~\ref{fig:basic-processes}(a) can be neglected.  In this case one
can assume that the numbers of the right- and left-moving fermions are
conserved, and the relaxation in the system is due to the processes of
Fig.~\ref{fig:basic-processes}(b) and (c).  A small temperature
gradient $\partial_xT$ still results in a dissipative contribution to
the energy current $-\kappa_{\rm ex}\partial_xT$ proportional to it,
but with a different thermal conductivity $\kappa_{\rm ex}$.  The
transport coefficient $\kappa_{\rm ex}$ was recently introduced in the
two-fluid hydrodynamic theory of one-dimensional quantum liquids
\cite{matveev_propagation_2018}.  It describes the thermal
conductivity of the gas of elementary excitations of the quantum
liquid and appears in the expressions for damping of the sound modes
in this system.

In this paper we develop a microscopic theory of thermal conductivity
of a one-dimensional Fermi gas with weak interactions between the
particles.  Our main focus is on the case of spinless fermions, for
which the relaxation processes have been studied in considerable
detail \cite{khodas_fermi-luttinger_2007, matveev_decay_2013,
  matveev_equilibration_2012, ristivojevic_relaxation_2013}.  At $T\ll
\mu$ our result for the thermal conductivity $\kappa$ is consistent
with the phenomenological expression (\ref{eq:kappa_Luttinger}), while
also providing an expression for the relaxation time $\tau$ in terms
of the microscopic interaction potential.  More importantly, our
approach enables us to obtain the thermal conductivity of the gas of
excitations $\kappa_{\rm ex}$, for which no phenomenological theory is
available.  Because the relaxation processes are sensitive to the form
of interaction between fermions \cite{khodas_fermi-luttinger_2007,
  ristivojevic_relaxation_2013}, we find very different temperature
dependence of $\kappa_{\rm ex}$ for the short-range and Coulomb
interactions.

The paper is organized as follows.  In Sec.~\ref{sec:kappa} we use
Boltzmann equation approach to obtain a microscopic expression for the
thermal conductivity $\kappa$ of the degenerate one-dimensional Fermi
gas.  The same technique is applied to the calculation of the thermal
conductivity of the gas of excitations in Sec.~\ref{sec:kappa_ex},
where a general expression and the order of magnitude extimate of
$\kappa_{\rm ex}$ are obtained.  A careful evaluation of
$\kappa_{\rm ex}$ involves a detailed treatment of the relaxation
processes shown in Fig.~\ref{fig:basic-processes}(b) and (c), which is
presented in Sec.~\ref{sec:relaxation}.  We discuss our results in
Sec.~\ref{sec:discussion}.

\section{Thermal conductivity of the Fermi gas}
\label{sec:kappa}

\subsection{Boltzmann equation approach}
\label{sec:Boltzmann}

We start by evaluating the thermal conductivity of the one-dimensional
gas of spinless fermions with the energy spectrum
\begin{equation}
  \label{eq:varepsilon}
  \varepsilon_p=\frac{p^2}{2m},
\end{equation}
where $p$ is the momentum of the fermion and $m$ is its mass.  We will
subject the system to an infinitesimal temperature gradient
$\partial_x T$ and obtain the occupation numbers of the fermionic
states $n_p$ from the Boltzmann equation \cite{lifshitz_physical_1981}
\begin{equation}
  \label{eq:BoltzmannEq}
  \partial_tn_p+\frac{p}{m}\partial_xn_p
  =I[n_p].
\end{equation}
Weak interactions between fermions give rise to the scattering
processes accounted for by the collision integral $I[n_p]$.  In the
left-hand side of the Boltzmann equation interactions will be
neglected.  In this approximation the energy of the fermion
(\ref{eq:varepsilon}) does not depend on its position $x$, which
enabled us to omit an additional term
$-(\partial_x\varepsilon_p)\partial_pn_p$ in the left-hand side of
Eq.~(\ref{eq:BoltzmannEq}).

We are considering a translation-invariant system, in which collisions
between the fermions conserve not only the total number of particles
and energy of the system, but also its momentum.  In this case even in
thermodynamic equilibrium the system can move with respect to the lab
frame with some velocity $u$, and the equilibrium occupation numbers
of the fermionic states are given by
\begin{equation}
  \label{eq:equilibrium_distribution}
  n_p^{(0)}=\frac{1}{e^{\beta\varepsilon_p-\gamma p-\alpha}+1},
\end{equation}
where $\beta=1/T$, $\alpha=\beta\mu$, and $\gamma=\beta u$.  In the
presence of the temperature gradient $\partial_x T$ the occupation
numbers deviate from the equilibrium form
(\ref{eq:equilibrium_distribution}),
\begin{equation}
  \label{eq:distribution_with_correction}
  n_p=n_p^{(0)}+\delta n_p,
\end{equation}
where the small non-equilibrium correction $\delta
n_p\propto \partial_x T$.  The distribution function $n_p$ depends on
the spatial coordinate $x$.  We assume that the parameters
$\alpha(x)$, $\beta(x)$, and $\gamma(x)$ of the equilibrium part
$n_p^{(0)}$ of the distribution function
(\ref{eq:distribution_with_correction}) are chosen in such a way that
the particle, energy, and momentum densities of the system can be
evaluated by substituting $n_p^{(0)}$ for $n_p^{}$.  In other words,
we impose the conditions
\begin{equation}
  \label{eq:delta_n_p_conditions}
  \int\! \delta n_p\, dp=0,
\quad
  \int\! \varepsilon_p\delta n_p\, dp=0,
\quad
  \int\! p\,\delta n_p\, dp=0
\end{equation}
upon $\delta n_p$.

Our immediate goal is to obtain the thermal conductivity $\kappa$,
defined by the relation
\begin{equation}
  \label{eq:kappa_definition}
  j_Q=-\kappa \partial_x T.
\end{equation}
Here 
\begin{equation}
  \label{eq:heat_current_definition}
  j_Q=\int\frac{dp}{2\pi\hbar}\varepsilon_p\frac{p}{m}\delta n_p
\end{equation}
is the dissipative part of the energy current
\cite{lifshitz_physical_1981}.

We will obtain the correction $\delta n_p$ to the equilibrium
distribution function by solving the Boltzmann equation
(\ref{eq:BoltzmannEq}) written in the form
\begin{equation}
  \label{eq:BoltzmannEq-2}
  \dot n_p = I[n_p^{(0)}+\delta n_p],
\end{equation}
where $\dot n_p$ is given by the left-hand side of
Eq.~(\ref{eq:BoltzmannEq}).  Since the collision integral evaluated
for the equilibrium distribution $I[n_p^{(0)}]=0$, it is important to
keep the infinitesimal correction $\delta n_p$ in the right-hand side
of Eq.~(\ref{eq:BoltzmannEq-2}).  On the other hand, $\dot n_p$ should
be evaluated for $\delta n_p=0$, i.e.,
\begin{equation}
  \label{eq:dot_n_p}
   \dot n_p=\partial_tn_p^{(0)}+\frac{p}{m}\partial_xn_p^{(0)}.
\end{equation}
The thermal conductivity is defined in the steady-state regime, in
which the parameters $\alpha$, $\beta$, and $\gamma$ of the
equilibrium distribution function (\ref{eq:equilibrium_distribution})
do not depend on time.  In addition, the system is assumed to be
stationary, i.e., $\gamma=0$.  The latter condition is satisfied due
to the gradient of the chemical potential emerging in the system,
which compensates for the force acting on the system as a result of
the applied temperature gradient.  Thus we substitute into
Eq.~(\ref{eq:BoltzmannEq-2}) the equilibrium distribution
(\ref{eq:equilibrium_distribution}) with the parameters
\begin{equation}
  \label{eq:parameters}
  \alpha=\alpha(x),
\quad
  \beta=\beta(x),
\quad
  \gamma=0.
\end{equation}
This results in
\begin{equation}
  \label{eq:ndot}
  \dot n_p=-g_p^2[(\partial_x\beta)\varepsilon_p 
                -\partial_x\alpha]\frac{p}{m},
\end{equation}
where we have introduced
\begin{equation}
  \label{eq:g_p}
  g_p=\sqrt{n_p^{(0)}\Big(1-n_p^{(0)}\Big)}
     =\frac{1}{2\cosh\frac{\beta\varepsilon_p-\alpha}{2}}.
\end{equation}
The value of $\partial_x\alpha$ in Eq.~(\ref{eq:ndot}) can be found by
noticing that the scattering processes accounted for by the collision
integral $I[n_p]$ conserve the total momentum of the system, i.e.,
\begin{equation}
  \label{eq:momentum_conservation}
  \int \frac{dp}{2\pi\hbar}\, p\:\! \dot n_p=0.
\end{equation}
Imposing this condition on Eq.~(\ref{eq:ndot}), we obtain
\begin{equation}
  \label{eq:ndot_full}
  \dot n_p=-g_p^2\frac{\partial_x\beta}{m}[\varepsilon_p 
                -\widetilde\mu]p,
\quad
  \widetilde \mu = \frac{1}{2m}\frac{I_4}{I_2},
\end{equation}
where 
\begin{equation}
  \label{eq:I_lambda}
  I_\lambda=\int_{0}^{+\infty}dp\,g_p^2 p^\lambda.
\end{equation}
In the zero temperature limit $\widetilde \mu\to\mu$.

\subsection{Linearized collision integral}
\label{sec:linearized_collision_int}

Our next step is to obtain $\delta n_p$ by solving
Eq.~(\ref{eq:BoltzmannEq-2}).  Since we are interested in the linear
response to an infinitesimal temperature gradient, we can linearize
the collision integral in the right-hand side of
Eq.~(\ref{eq:BoltzmannEq-2}).  In addition, it is convenient to write
the resulting integral equation in terms of function $\phi_p$ defined
by
\begin{equation}
  \label{eq:x_p_definition}
  \delta n_p=g_p\phi_p.
\end{equation}
As a result, Eq.~(\ref{eq:BoltzmannEq-2}) becomes a linear integral
equation
\begin{equation}
  \label{eq:BoltzmannEq-3}
  \int \frac{dp'}{2\pi\hbar}\,K(p,p')\phi_{p'}=\frac{\dot n_p}{g_p}
\end{equation}
with a real symmetric kernel $K(p,p')$.  The latter property means
that one can, in principle, obtain an orthonormal set of
eigenfunctions of this integral operator,
\begin{equation}
  \label{eq:eigenvalue_problem}
  \int \frac{dp'}{2\pi\hbar}\,K(p,p')\phi_{p'}^{(l)} 
     =-\frac{1}{\tau_l^{}} \phi_p^{(l)},
\quad \langle\phi_p^{(l)}|\phi_p^{(l')}\rangle=\delta_{l,l'},
\end{equation}
with real eigenvalues $-1/\tau_l^{}$.  Here we defined the inner product
by
\begin{equation}
  \label{eq:inner_product}
  \langle\phi_p|\psi_p\rangle = \int \frac{dp}{2\pi\hbar}\,\phi_p\psi_p.
\end{equation}
Because the collisions result in the evolution of the distribution
function toward equilibrium, the relaxation rates $\tau_l^{-1}$ are
non-negative.

The full set of eigenfunctions $\phi_p^{(l)}$ includes three modes
with zero eigenvalues.  The existence of such zero modes is due to the
conservation of the total number of particles $N$, energy $E$, and
momentum $P$ of the system.  They are obtained by small variations of
parameters $\alpha$, $\beta$, and $\gamma$ in the expression for the
equilibrium distribution (\ref{eq:equilibrium_distribution}),
\begin{equation}
  \label{eq:zero_modes}
  \phi_p^{(N)}=g_p,
\quad
  \phi_p^{(E)}=\left(\varepsilon_p- \frac{I_2}{2mI_0}\right)g_p,
\quad
  \phi_p^{(P)}=p\;\! g_p,
\end{equation}
where we orthogonalized the modes, but omitted the normalization
constants.  Indeed, any such variation transforms $n_p^{(0)}$ to
another equilibrium distribution.  Collisions do not modify
equilibrium distributions, resulting in vanishing eigenvalues in
Eq.~(\ref{eq:eigenvalue_problem}).

We can now expand $\phi_p$ in the basis of the eigenfunctions
$\phi_p^{(l)}$ and obtain a formal solution of the integral equation
(\ref{eq:BoltzmannEq-3}),
\begin{equation}
  \label{eq:phi_p_series}
  \phi_p=-\frac{\partial_x T}{mT^2}
          \sum_{l\neq N,E,P}\tau_l^{} 
          \langle \phi_p^{(l)}|
          g_p(\varepsilon_p-\widetilde\mu)p\rangle
          \phi_p^{(l)},
\end{equation}
where we used Eq.~(\ref{eq:ndot_full}) for $\dot n_p$.  In the sum in
Eq.~(\ref{eq:phi_p_series}) we excluded the zero modes
(\ref{eq:zero_modes}) for which the corresponding $\tau_l$ is
infinite.  This is due to the fact that the overlaps of $
g_p(\varepsilon_p-\widetilde\mu)p$ with $\phi_p^{(N)}$ and
$\phi_p^{(E)}$ vanish due to opposite symmetries with respect to
$p\to-p$, while the overlap with $\phi_p^{(P)}$ vanishes because of
the momentum conservation condition (\ref{eq:momentum_conservation}).
As a result, the conditions (\ref{eq:delta_n_p_conditions}) for
$\delta n_p$ are satisfied.

Next, we notice that due to the last of the conditions
(\ref{eq:delta_n_p_conditions}) one can replace $\varepsilon_p\to
\varepsilon_p -\widetilde\mu$ in the expression for the dissipative
contribution (\ref{eq:heat_current_definition}) to the energy current,
after which the latter becomes
\begin{equation}
  \label{eq:heat_current_algebraic}
  j_Q=\frac{1}{m}\langle g_p(\varepsilon_p -\widetilde\mu)p|\phi_p\rangle.
\end{equation}
Substitution of Eq.~(\ref{eq:phi_p_series}) and comparison with
Eq.~(\ref{eq:kappa_definition}) yield
\begin{equation}
  \label{eq:kappa_general}
  \kappa=\frac{1}{m^2T^2}
       \sum_{l\neq N,E,P}\tau_l^{}\,
          \langle \phi_p^{(l)}|
          g_p(\varepsilon_p-\widetilde\mu)p\rangle^2.
\end{equation}
This expression gives the thermal conductivity of the one-dimensional
spinless Fermi gas at any temperature.  Significant further progress
in understanding thermal conductivity can be made in the regime of low
temperature, $T\ll \mu$.

\subsection{Thermal conductivity at low temperatures}
\label{sec:kappa_low_T}

At low temperatures the relaxation of the one-dimensional Fermi gas is
dominated by the processes shown in Fig.~\ref{fig:basic-processes}.
The process of Fig.~\ref{fig:basic-processes}(a) is exponentially
suppressed, whereas relaxation rate $\tau_{\rm ex}^{-1}$ associated
with the processes of Fig.~\ref{fig:basic-processes}(b) and (c) has a
power-law temperature dependence.  It is important to note, however,
that understanding the full relaxation of the system to equilibrium
requires accounting for the effect of the processes of
Fig.~\ref{fig:basic-processes}(a), because the remaining processes do
not change the numbers of the left- and right-moving particles in the
system.

Let us consider a Fermi gas with zero total momentum, which relaxes to
the equilibrium distribution (\ref{eq:equilibrium_distribution}) with
$\gamma=0$.  Keeping in mind the presence of two very different
relaxation times $\tau\gg \tau_{\rm ex}$, one concludes that the
relaxation of the degenerate one-dimensional Fermi gas proceeds in two
steps.  First, at the time scales of the order of $\tau_{\rm ex}$ the
particle-hole excitations come to equilibrium with each other, but the
chemical potentials of the left- and right-moving particles remain
different.  This means that the distribution function takes the form
\begin{equation}
  \label{eq:n_p_partial}
  n_p=\frac{\theta(p)}
                    {e^{\beta\varepsilon_p-\gamma p-\alpha-\delta\alpha}+1}
                  +\frac{\theta(-p)}
                    {e^{\beta\varepsilon_p-\gamma p-\alpha+\delta\alpha}+1}.
\end{equation}
The presence of the new parameter $\delta\alpha$ in the distribution
function accounts for the fact that in addition to the total number of
particles, energy, and momentum of the system, the difference $J$ of
the numbers of the right- and left-moving particles is also conserved
at time scales $t\ll\tau$.

To linear order in $\delta\alpha$ and $\gamma$, the momentum density of
the Fermi gas with the distribution (\ref{eq:n_p_partial}) is
$(\delta\alpha I_1+\gamma I_2)/\pi\hbar$.  Thus the parameter
$\gamma$ in Eq.~(\ref{eq:n_p_partial}) is related to $\delta \alpha$
by $\gamma =-(I_1/I_2)\delta\alpha$, and the deviation of the
distribution function (\ref{eq:n_p_partial}) from the equilibrium form
$(e^{\beta\varepsilon_p-\alpha}+1)^{-1}$ is
\begin{equation}
  \label{eq:delta_n_p_partial}
  \delta n_p=\delta\alpha\, g_p^2
              \left({\rm sgn\,}p-\frac{I_1}{I_2}p\,\right).
\end{equation}
At the second stage of the relaxation process the system slowly
approaches equilibrium, $\delta\alpha\propto e^{-t/\tau}$, while the
momentum dependence of $\delta n_p$ retains the form
(\ref{eq:delta_n_p_partial}).  We therefore conclude that the
eigenvalue problem (\ref{eq:eigenvalue_problem}) for the collision
integral has the solution
\begin{equation}
  \label{eq:phi^J}
  \phi^{(J)}_p=C_J\,g_p
              \left({\rm sgn\,}p-\frac{I_1}{I_2}p\,\right)
\end{equation}
with exponentially small eigenvalue $-1/\tau$.  The
constant
\begin{equation}
  \label{eq:C_J}
  C_J=\sqrt{\frac{\pi\hbar I_2}{I_2 I_0 -I_1^2}}
\end{equation}
is obtained from the normalization condition in
Eq.~(\ref{eq:eigenvalue_problem}).

The remaining relaxation modes $\phi_p^{(l)}$ describe how the
non-equilibrium distribution function approaches the form
(\ref{eq:n_p_partial}) at the first stage of the relaxation process,
dominated by the processes of Fig.~\ref{fig:basic-processes}(b) and (c).
The corresponding relaxation times $\tau_l\sim\tau_{\rm ex}$ are much
smaller than $\tau\propto e^{\mu/T}$.  Thus, at low temperatures $T\ll
\mu$, the thermal conductivity (\ref{eq:kappa_general}) is dominated
by the mode (\ref{eq:phi^J}).  Evaluation of the corresponding matrix
element is straightforward,
\begin{equation}
  \label{eq:phi^J_matrix_element}
  \langle \phi_p^{(J)}|g_p(\varepsilon_p-\widetilde\mu)p\rangle
  =\frac{C_J}{2\pi\hbar m}\,\frac{I_3 I_2 -I_4I_1}{I_2}.
\end{equation}
At $T\ll\mu$ the integrals (\ref{eq:I_lambda}) can be approximated by
the Sommerfeld expansion
\begin{eqnarray}
  \label{eq:I_lambda_approx}
  I_\lambda&\simeq& mT(2m\mu)^{\frac{\lambda-1}{2}}
          \bigg(
           1+\frac{(\lambda-1)(\lambda-3)}{24}\frac{\pi^2T^2}{\mu^2}
\nonumber\\
          &&+\frac{7(\lambda-1)(\lambda-3)(\lambda-5)(\lambda-7)}{5760}
             \frac{\pi^4T^4}{\mu^4}
          \bigg).
\end{eqnarray}
For the dominant contribution to the thermal conductivity
(\ref{eq:kappa_general}) we then find
\begin{equation}
  \label{eq:kappa_low_T}
  \kappa\simeq\frac{\tau}{m^2T^2}
          \langle \phi_p^{(J)}|
          g_p(\varepsilon_p-\widetilde\mu)p\rangle^2
        \simeq\frac{\pi T\tau}{3\hbar}\sqrt{\frac{2\mu}{m}}.
\end{equation}
The speed $v$ of the low-energy elementary excitations in a Fermi gas
at $T\ll \mu$ is the Fermi velocity $v_F=\sqrt{2\mu/m}$.  Therefore
the result (\ref{eq:kappa_low_T}) is consistent with the
phenomenological expression (\ref{eq:kappa_Luttinger}) for the thermal
conductivity of a spinless one-dimensional quantum liquid.  We note
that both Eqs.~(\ref{eq:kappa_Luttinger}) and (\ref{eq:kappa_low_T})
express $\kappa$ in terms of the relaxation time $\tau$.  The latter
has been studied in some detail phenomenologically in
Refs.~\cite{matveev_equilibration_2012, matveev_scattering_2012},
where it was expressed in terms of the quasiparticle spectrum of the
quantum liquid.  In Sec.~\ref{sec:discussion} we discuss how one can
obtain a microscopic expression for $\tau$ in terms of interactions
between fermions.

\section{Thermal conductivity of the gas of elementary excitations}
\label{sec:kappa_ex}

In Sec.~\ref{sec:kappa} we found that at $T\ll\mu$ the thermal
conductivity of the Fermi gas (\ref{eq:kappa_low_T}) is proportional
to the relaxation time $\tau$ and is, therefore, exponentially large.
This result holds as long as thermal conductivity is measured at
frequencies $\omega\ll\tau^{-1}$.  On the other hand, interesting new
behavior of one-dimensional systems is expected in the broad range of
frequencies
\begin{equation}
  \label{eq:frequency_range} \tau^{-1} \ll \omega \ll \tau_{\rm
ex}^{-1}.
\end{equation}
It was shown recently \cite{matveev_second_2017, matveev_hybrid_2018}
that in this regime one-dimensional systems behave like superfluids
and support two sound modes, in contrast to a single sound mode at
$\omega\ll\tau^{-1}$.  

In the presence of an ordinary sound wave the temperature of the
system depends on position.  This results in dissipation, which is
proportional to the thermal conductivity $\kappa$ and contributes to
the attenuation of sound \cite{landau_fluid_2013}.  The same physics
applies in the two-sound regime (\ref{eq:frequency_range}), but the
resulting contribution to the sound attenuation is controlled by a
different thermal transport coefficient $\kappa_{\rm ex}$
\cite{matveev_propagation_2018}, which has the meaning of the
dissipative part of the thermal conductivity at frequencies in the
range (\ref{eq:frequency_range}).

To evaluate $\kappa_{\rm ex}$ we will assume that the exponentially
long relaxation time $\tau\to\infty$ and then take the limit
$\omega\to0$.  In other words, we will find the thermal conductivity
assuming that the relaxation processes conserve not only the number of
particles, energy, and momentum of the system, but also the difference
$J$ of the numbers of right- and left-moving particles.  We now adapt
the evaluation of the thermal conductivity in Sec.~\ref{sec:Boltzmann}
and \ref{sec:linearized_collision_int} to account for the this
additional conservation law.

First of all, the equilibrium distribution $n_p^{(0)}$ now takes the
form (\ref{eq:n_p_partial}) and depends on four parameters, $\alpha$,
$\beta$, $\gamma$, and $\delta\alpha$.  A small temperature gradient
$\partial_x T$ results in a small deviation of the distribution
function from the equilibrium form, see
Eq.~(\ref{eq:distribution_with_correction}).  Because of the fourth
conservation law, in addition to Eq.~(\ref{eq:delta_n_p_conditions})
we impose the condition
\begin{equation}
  \label{eq:delta_n_p_fourth_condition}
    \int\! {\rm sgn\,}p\,\delta n_p\, dp=0
\end{equation}
to ensure that the value of $J$ is determined by the equilibrium
part of $n_p$. 

We next obtain $\delta n_p$ in the expression for the dissipative
energy current (\ref{eq:heat_current_definition}) by solving the
Boltzmann equation in the form (\ref{eq:BoltzmannEq-2}) with $\dot
n_p$ given by Eq.~(\ref{eq:dot_n_p}).  The conditions
(\ref{eq:parameters}) must be modified to account for the extra
conservation law.  First, the condition of zero total momentum of the
system now takes the form $\gamma =-(I_1/I_2)\delta\alpha$, see
Sec.~\ref{sec:kappa_low_T}.  Second, in the absence of the relaxation
processes of Fig.~\ref{fig:basic-processes}(a) one can no longer
exclude the possibility of a time-dependent difference of the chemical
potentials of the right- and left-moving particles.  This results in
the following assumptions regarding the parameters of the equilibrium
distribution (\ref{eq:n_p_partial})
\begin{equation}
  \label{eq:parameters-2}
  \alpha=\alpha(x),
\quad
  \beta=\beta(x),
\quad
  \gamma=-\frac{I_1}{I_2}\,\delta\alpha,
\quad
  \delta\alpha=\delta\alpha(t).
\end{equation}
Substituting the expression (\ref{eq:n_p_partial}) for $n_p^{(0)}$ in
Eq.~(\ref{eq:dot_n_p}) we get
\begin{equation}
  \label{eq:ndot-2}
  \dot n_p=-g_p^2\bigg\{[(\partial_x\beta)\varepsilon_p 
                -\partial_x\alpha]\frac{p}{m}
           - (\partial_t\delta\alpha)
            \bigg({\rm sgn\,}p-\frac{I_1}{I_2}p\,\bigg)\bigg\}
\end{equation}
The values of $\partial_x\alpha$ and $\partial_t\delta\alpha$ are
determined by imposing the condition (\ref{eq:momentum_conservation})
of conservation of momentum in collisions along with the new condition
of conservation of $J$,
\begin{equation}
  \label{eq:J_conservation}
  \int \frac{dp}{2\pi\hbar}\, {\rm sgn\,}p\, \dot n_p=0.
\end{equation}
This yields
\begin{equation}
  \label{eq:ndot_full-2}
  \dot n_p=-g_p^2\frac{\partial_x\beta}{2m^2}
           \nu_p,
\end{equation}
where
\begin{equation}
  \label{eq:nu_p}
  \nu_p= p^3-\frac{I_4I_0 - I_3I_1}{I_2I_0 - I_1^2}p
                 +\frac{I_4I_1 - I_3I_2}{I_2I_0 - I_1^2}{\rm sgn\,}p.
\end{equation}
We note that by imposing the conservation laws
(\ref{eq:momentum_conservation}) and (\ref{eq:J_conservation}) we
ensured that the function $\phi_p=g_p\nu_p$ is orthogonal to both the
usual zero modes (\ref{eq:zero_modes}) and the additional zero mode
(\ref{eq:phi^J}) corresponding to the conservation of $J$.

The next step is to find the non-equilibrium correction $\delta n_p$
to the distribution function by solving the integral equation
(\ref{eq:BoltzmannEq-3}).  Writing $\delta n_p$ as prescribed in
Eq.~(\ref{eq:x_p_definition}), and expanding $\phi_p$ in the basis of
the eigenfunctions $\phi_p^{(l)}$ of the linearized collision integral
with non-zero eigenvalues, we obtain
\begin{equation}
  \label{eq:phi_p_series-2}
  \phi_p=-\frac{\partial_x T}{2m^2T^2}
          \sum_{l\neq N,E,P,J}\tau_l^{} 
          \langle \phi_p^{(l)}| g_p\nu_p\rangle
          \phi_p^{(l)}.
\end{equation}
Because $\phi_p$ is orthogonal to the four zero modes
(\ref{eq:zero_modes}) and (\ref{eq:phi^J}), the conditions
(\ref{eq:delta_n_p_conditions}) and
(\ref{eq:delta_n_p_fourth_condition}) imposed on $\delta n_p$ are
satisfied.

Using Eq.~(\ref{eq:delta_n_p_fourth_condition}) and the last of the
conditions (\ref{eq:delta_n_p_conditions}), it is convenient to
replace $\varepsilon_p p\to \nu_p/2m$ in the definition
(\ref{eq:heat_current_definition}) of $j_Q$, which results
in
\begin{equation}
  \label{eq:heat_current_algebraic-2}
  j_Q=\frac{1}{2m^2}\langle g_p\nu_p|\phi_p\rangle.
\end{equation}
Substitution of the expression (\ref{eq:phi_p_series-2}) for $\phi_p$
yields $j_Q=-\kappa_{\rm ex}\partial_x T$ with
\begin{equation}
  \label{eq:kappa_ex_general}
  \kappa_{\rm ex}=\frac{1}{4m^4T^2}
       \sum_{l\neq N,E,P,J}\tau_l^{}\,
          \langle \phi_p^{(l)}|
          g_p\nu_p\rangle^2.
\end{equation}
Unlike the similar expression (\ref{eq:kappa_general}) for $\kappa$,
the result (\ref{eq:kappa_ex_general}) assumes the low-temperature
regime, $T\ll \mu$, because the transport coefficient $\kappa_{\rm ex}$
is defined only at $\tau\gg \tau_{\rm ex}$.

At $T\ll \mu$ the particle and hole excitations are confined to the
vicinities of the two Fermi points $p=\pm p_F$, where
$p_F=\sqrt{2m\mu}$.  For such values of momentum we can use
Eq.~(\ref{eq:I_lambda_approx}) to approximate $\nu_p$ defined by
Eq.~(\ref{eq:nu_p}) as
\begin{equation}
  \label{eq:nu_p_approx}
  \nu_p=3p_F\bigg[(|p|-p_F)^2-\frac{\pi^2 T^2}{3v_F^2}\bigg]{\rm sgn\,}p.
\end{equation}
For typical values of momentum, $|p|-p_F\sim T/v_F$, the leading order
correction to Eq.~(\ref{eq:nu_p_approx}) scales as $T^3$ and can be
easily shown to give a subleading contribution to
Eq.~(\ref{eq:kappa_ex_general}).  Thus, the dominant contribution to
$\kappa_{\rm ex}$ can be obtained by combining
Eqs.~(\ref{eq:kappa_ex_general}) and (\ref{eq:nu_p_approx}).

We now obtain an order of magnitude estimate of the transport
coefficient $\kappa_{\rm ex}$ using Eqs.~(\ref{eq:kappa_ex_general})
and (\ref{eq:nu_p_approx}).  The typical relaxation time
$\tau_l^{}\sim\tau_{\rm ex}$.  Evaluation of the inner product
(\ref{eq:inner_product}) adds a factor of order $T/\hbar v_F$.  Thus
the normalization of the eigenfunctions prescribed by
Eq.~(\ref{eq:eigenvalue_problem}) gives $\phi_p^{(l)}\sim \sqrt{\hbar
  v_F/T}$.  Combining these estimates we find
\begin{equation}
  \label{eq:kappa_ex_estimate}
  \kappa_{\rm ex}\sim\frac{T^3 v_F\tau_{\rm ex}}{\hbar\mu^2}.
\end{equation}
To evaluate $\tau_{\rm ex}$ and obtain the numerical prefactor in
Eq.~(\ref{eq:kappa_ex_estimate}), one has to carefully consider the
collision integral of the Boltzmann equation (\ref{eq:BoltzmannEq}).
We present this treatment in the next section.

\section{Relaxation of the degenerate Fermi gas to equilibrium}
\label{sec:relaxation}

We now evaluate the transport coefficient $\kappa_{\rm ex}$ in terms
of the two-particle interaction potential $U(x)$ between the fermions.
Our prescription (\ref{eq:kappa_ex_general}) requires one to find the
full spectrum of the relaxation rates $\tau_l^{-1}$ in the system as
well as the respective relaxation modes $ \phi_p^{(l)}$ by solving the
eigenvalue problem (\ref{eq:eigenvalue_problem}) for the linearized
collision integral.  We will show that the problem simplifies
considerably for the interaction potentials that decay slowly with the
distance between particles, such as the Coulomb interaction.  In this
case one of the relaxation modes coincides with $\nu_p$ given by
Eq.~(\ref{eq:nu_p_approx}) up to a normalization factor.  As a result
the sum in Eq.~(\ref{eq:kappa_ex_general}) includes just one term, and
the evaluation of $\kappa_{\rm ex}$ simplifies considerably.  This is
not the case for interaction potentials that decay rapidly with the
distance between fermions, which will be considered separately.

\subsection{Coulomb and dipole-dipole interactions}
\label{sec:Coulomb}

In the case of charged particles, their interactions are usually
dominated by Coulomb repulsion.  For particles with charge $e$
confined to a narrow channel the interaction potential takes the form
$U(x)={e^2}/{|x|}$ at distances $x$ that are large compared to the the
width of the channel $w$.  The behavior of $U(x)$ at $x\lesssim w$ is
determined by the nature of the confining potential.  The study of the
relaxation spectrum requires evaluation of the Fourier transform of
the interaction potential
\begin{align}
\label{eq:Fourier}
V(p)=\int_{-\infty}^{+\infty}  U(x) e^{i p x/\hbar} dx.
\end{align}
Substitution of $U(x)={e^2}/{|x|}$ into Eq.~(\ref{eq:Fourier}) results
in a logarithmic singularity.  We therefore account properly for the
short-distance behavior of $U(x)$ for particles confined to a channel
of width $w$, see Appendix~\ref{sec:Fourier}.  At low momenta, $|p|\ll
\hbar/w$, we find
\begin{align}
\label{eq:Coulomb}
V(p)=2e^2 \ln\left(\frac{\hbar}{|p|w} \right)\left(1+\frac{w^2p^2}{\hbar^2}\right).
\end{align}
A numerical factor in the argument of the logarithm in
Eq.~(\ref{eq:Coulomb}) depends on the details of the confinement and
is not included in the above expression.

Another important special case is dipole-dipole interaction
$U(x)={\Upsilon}/{|x|^3}$.  It can be realized, for example, in a
quantum wire in the vicinity of a metal gate parallel to it.  In this
case $\Upsilon=2e^2d^2$, where $d$ is the distance between the wire
and the gate. The Fourier transform of dipole-dipole interaction is
\begin{align}
\label{eq:dipole-dipole}
V(p)=-\Upsilon\frac{p^2}{\hbar^2}\ln\left(\frac{\hbar}{|p| w}\right).
\end{align}
Similarly to Eq.~(\ref{eq:Coulomb}) for Coulomb interaction,
Eq.~(\ref{eq:dipole-dipole}) is written within the logarithmic
accuracy and is restricted to small momenta, $|p|\ll \hbar/w$.  In
Eq.~(\ref{eq:dipole-dipole}) we omitted a large constant term that
corresponds to the contact interaction. For spinless fermions, Pauli
principle forbids two particles to occupy the same position in space,
and thus the contact interaction does not affect this system.

A special feature of slowly decaying potentials, such as Coulomb and
dipole-dipole ones, is that the relaxation processes which involve
co-propagating particles, shown in Fig.~\ref{fig:basic-processes}(c),
occur at a higher rate than the ones that involve counter-propagating
particles, see Fig~\ref{fig:basic-processes}(b). The estimate of the
corresponding rates can be obtained from the results of
Ref.~\cite{ristivojevic_relaxation_2013}, which studied the relaxation
of quasiparticles with energies much greater than $T$.  The typical
decay rate of thermal quasiparticles is $\tau_{\rm c}^{-1}\propto T^2$
for Coulomb interaction (\ref{eq:Coulomb}), and $\tau_{\rm
  c}^{-1}\propto T^6$ for dipole-dipole interaction
(\ref{eq:dipole-dipole}).  These decay rates are larger than the ones
involving processes depicted in Fig.~\ref{fig:basic-processes}(b),
which occur at rates $\tau_{\rm b}^{-1}\propto T^3$ and $\tau_{\rm
  b}^{-1}\propto T^7$, respectively. As a result, at time scales
longer than $\tau_{\rm c}$ but shorter than $\tau_{\rm b}$, each
branch of excitations independently achieves equilibrium characterized
by its own parameters $\alpha$, $\beta$, and $\gamma$. Thus, the
distribution function takes a partially-equilibrated form
\cite{micklitz_transport_2010}
\begin{align}
\label{eq:n_p_partial30}
n_p={}&\frac{\theta(p)}{e^{\beta_R \varepsilon_p-\gamma_R p-\alpha_R}+1}+\frac{\theta(-p)}{e^{\beta_L \varepsilon_p-\gamma_L p-\alpha_L}+1}.
\end{align}
Further relaxation of the distribution (\ref{eq:n_p_partial30}) toward
the form (\ref{eq:n_p_partial}) is controlled by the processes shown
in Fig.~\ref{fig:basic-processes}(b), with the relaxation time
$\tau_{\rm b}\gg \tau_{\rm c}$.

In the following we will calculate the eigenmodes and the
corresponding rates for relaxation of the distribution
(\ref{eq:n_p_partial30}), which will be sufficient to obtain the
thermal conductivity (\ref{eq:kappa_ex_general}).  We note that this
is a much simpler problem than the full solution of the eigenvalue
problem (\ref{eq:eigenvalue_problem}). Instead of diagonalizing the
full collision integral, which has an infinite number of eigenmodes,
the problem is reduced to the study of the evolution of only six
parameters in Eq.~(\ref{eq:n_p_partial30}).

At small deviations from equilibrium, we expand
Eq.~(\ref{eq:n_p_partial30}) as $n_p=\left(e^{\beta
    \varepsilon_p-\alpha}+1\right)^{-1}+\delta n_p$, where
\begin{align}
\label{eq:correction-n_p}
\delta n_p{}&=\theta(p)g_p^2 \left[-(\beta_R-\beta)\varepsilon_p+\gamma_R p+\alpha_R-\alpha \right]\notag\\
&+\theta(-p)g_p^2 \left[-(\beta_L-\beta)\varepsilon_p +\gamma_L p+\alpha_L-\alpha \right]. 
\end{align}
The relaxation of $\delta n_p$ of Eq.~(\ref{eq:correction-n_p}) is
constrained by the four conditions given by
Eqs.~(\ref{eq:delta_n_p_conditions}) and
(\ref{eq:delta_n_p_fourth_condition}). This leads to four equations
for the six parameters of Eq.~(\ref{eq:correction-n_p}). We use them
to express four parameters as a function of the two remaining ones,
which we select to be $\beta_R-\beta_L$ and $\gamma_R-\gamma_L$. The
correction to the distribution function (\ref{eq:correction-n_p}) then
takes the form
\begin{align}
\label{eq:delta_n_p_30}
\delta n_p=-\frac{g_p^2}{4m}(\beta_R-\beta_L) \eta_p -\frac{g_p^2}{4p_F}(\gamma_R-\gamma_L) \sigma_p ,
\end{align}
where
\begin{gather}
\label{eq:eta}
\eta_p=p^2\,\mathrm{sgn}\, p- \frac{I_3I_0-I_2I_1}{I_2I_0-I_1^2}p +\frac{I_3I_1-I_2^2}{I_2I_0-I_1^2}\textrm{sgn}\,p,\\
\label{eq:sigma}
\sigma_p=2p_F\left(\frac{I_3 I_0-I_2 I_1}{I_4 I_0-I_2^2}p^2-|p|+\frac{I_4 I_1-I_3 I_2}{I_4 I_0-I_2^2}\right).
\end{gather}
We note that $\eta_p$ is an odd function of $p$, whereas $\sigma_p$ is
an even one. At low temperature we find
\begin{gather}
\label{eq:etaapprox}
\eta_p=\left[(|p|-p_F)^2-\frac{\pi^2 T^2}{3v_F^2}\right]\mathrm{sgn}\,p,\\
\sigma_p=(|p|-p_F)^2-\frac{\pi^2 T^2}{3v_F^2}.
\end{gather}
In this regime $\eta_p$ and $\sigma_p$ are equal in absolute value. 

To find the evolution of the distribution function
(\ref{eq:n_p_partial30}), we consider the rate of change of the
occupation number of the state $p$ due to three-particle collisions
\begin{widetext}
\begin{eqnarray}
	\label{eq:dotn_p}
	\dot n_p&=&-
	\frac{1}{12}\sum_{\substack{p_1,p_2\\p',p_1',p_2'}}\frac{2\pi}{\hbar}
	\left|\mathcal{A}_{p,p_1,p_2}^{p',p_1',p_2'}\right|^2
	\delta(\varepsilon_p+\varepsilon_{p_1}+\varepsilon_{p_2}
	-\varepsilon_{p'}-\varepsilon_{p_1'}-\varepsilon_{p_2'})
	\nonumber\\
	&&\qquad\qquad\quad\times \left[n_pn_{p_1}n_{p_2}(1-n_{p'})(1-n_{p_1'})(1-n_{p_2'})
	-(1-n_{p})(1-n_{p_1})(1-n_{p_2})n_{p'}n_{p_1'}n_{p_2'}\right].
\end{eqnarray}
Here the scattering matrix element
$\mathcal{A}_{p,p_1,p_2}^{p',p_1',p_2'}$ depends on the details of the
two-body interaction potential \cite{imambekov_one-dimensional_2012,
  ristivojevic_relaxation_2013} and will be discussed below.  The
factor $1/12$ accounts for $12$ identical configurations that exist
due to unrestricted summations over the two initial ($p_1$ and $p_2$)
and three final ($p'$, $p_1'$, and $p_2'$) states.  

For systems close to the thermal equilibrium, we can linearize the
occupation factors by using
Eqs.~(\ref{eq:distribution_with_correction}) and
(\ref{eq:x_p_definition}). This yields
	\begin{align}
	\label{eq:BElin}
	\dot\phi_p=-\frac{1}{12 g_p} \sum_{\substack{p_1,p_2\\p',p_1',p_2'}}W_{p,p_1,p_2}^{p',p_1',p_2'} 
	\left(\frac{\phi_{p}}{g_{p}} + \frac{\phi_{p_1}}{g_{p_1}}+ \frac{\phi_{p_2}}{g_{p_2}}- \frac{\phi_{p'}}{g_{p'}}- \frac{\phi_{p_1'}}{g_{p_1'}}- \frac{\phi_{p_2'}}{g_{p_2'}}\right),
	\end{align}
	where we introduced
	\begin{align}
	\label{eq:W}
	W_{p,p_1,p_2}^{p',p_1',p_2'}=\frac{2\pi}{\hbar}\left|\mathcal{A}_{p,p_1,p_2}^{p',p_1',p_2'}\right|^2
	\delta(\varepsilon_p+\varepsilon_{p_1}+\varepsilon_{p_2}
	-\varepsilon_{p'}-\varepsilon_{p_1'}-\varepsilon_{p_2'})g_p g_{p_1} g_{p_2} g_{p'} g_{p_1'} g_{p_2'}.
	\end{align}
Using the separation of variables we transform Eq.~(\ref{eq:BElin})
into the eigenvalue problem [cf.~Eq.~(\ref{eq:eigenvalue_problem})]
\begin{align}
	\label{eq:BEdiag1}
	\frac{1}{\tau_l}\phi_p^{(l)}=\frac{1}{12 g_p} \sum_{\substack{p_1,p_2\\p',p_1',p_2'}}W_{p,p_1,p_2}^{p',p_1',p_2'}\left(
	\frac{\phi_{p}^{(l)}}{g_{p}} + \frac{\phi_{p_1}^{(l)}}{g_{p_1}}+ \frac{\phi_{p_2}^{(l)}}{g_{p_2}}- \frac{\phi_{p'}^{(l)}}{g_{p'}}- \frac{\phi_{p_1'}^{(l)}}{g_{p_1'}}- \frac{\phi_{p_2'}^{(l)}}{g_{p_2'}}\right).
\end{align}
Here the eigenvalue $1/\tau_l$ represents the relaxation rate
associated with the eigenfunction $\phi_p^{(l)}$ of the collision
operator.  Multiplying both sides in Eq.~(\ref{eq:BEdiag1}) by
$\phi_p^{(l)}$ and performing the summation over $p$, we find
	\begin{align}
	\label{eq:BEdiag}
	\frac{1}{\tau_l}\sum_p (\phi_p^{(l)})^2=\frac{1}{72} \sum_{\substack{p,p_1,p_2\\p',p_1',p_2'}}W_{p,p_1,p_2}^{p',p_1',p_2'}\left(
	\frac{\phi_{p}^{(l)}}{g_{p}} + \frac{\phi_{p_1}^{(l)}}{g_{p_1}}+ \frac{\phi_{p_2}^{(l)}}{g_{p_2}}- \frac{\phi_{p'}^{(l)}}{g_{p'}}- \frac{\phi_{p_1'}^{(l)}}{g_{p_1'}}- \frac{\phi_{p_2'}^{(l)}}{g_{p_2'}}\right)^2.
	\end{align}
\end{widetext}
Here we used the symmetries of $W$ to extend the summation over six
equivalent terms, thereby conveniently making the right-hand side
symmetric. As a result we have obtained an extra 1/6 prefactor and the
square of the expression in parentheses in the right-hand side. We
notice that Eqs.~(\ref{eq:dotn_p})--(\ref{eq:BEdiag}) do not have
restriction on the summation range and therefore apply for all three
kinds of processes shown in Fig.~\ref{fig:basic-processes}.

The momentum inversion symmetry of $W$ enables one to classify the
eigenfunctions of Eq.~(\ref{eq:BEdiag1}) with respect to parity. This
guarantees that the eigenfunctions in Eq.~(\ref{eq:BEdiag1}) will be
either odd or even in $p$.  Thus, the odd and even components of
$\phi_p$ obtained from Eq.~(\ref{eq:delta_n_p_30}) as $\delta
n_p/g_p$, which are proportional to $g_p\eta_p$ and even
$g_p\sigma_p$, respectively, relax as two independent eigenmodes.  We
now normalize them and introduce
\begin{gather}
\label{eq:phi^E}
\phi_p^{(\eta)}=C_\eta g_p \eta_p,\\
\label{eq:phi^P}
\phi_p^{(\sigma)}=C_\sigma g_p \sigma_p,
\end{gather}
where to leading order in $T/\mu$ the normalization constants are
equal,
\begin{align}
C_\eta=C_\sigma=\sqrt\frac{45\hbar v_F^5}{16\pi^3 T^5}.
\end{align}

Next, we calculate the relaxation rates corresponding to
$\phi_p^{(\eta)}$ and $\phi_p^{(\sigma)}$ using
Eq.~(\ref{eq:BEdiag}). We split the six-fold summation in that
expression into $2^6=64$ six-fold summations over momenta that are
either positive or negative. Out of 64 different terms, 44 describe
particle backscattering, i.e., the numbers of particles with positive
(negative) momenta are different in the initial and the final states
[see, for example, Fig.~\ref{fig:basic-processes}(a)].  At low
temperatures, such processes occur at exponentially long time scales
as discussed in Sec.~\ref{sec:kappa}. They determine the thermal
conductivity of the system $\kappa$ but must be neglected in
evaluating the thermal conductivity of the gas of excitations
$\kappa_{\rm ex}$.

Out of the remaining 20 terms, two have all the six momenta of the
same sign. The expression in parentheses in Eq.~(\ref{eq:BEdiag})
evaluated for such restricted range of momenta, i.e., all positive or
all negative, nullifies for the two eigenfunctions (\ref{eq:phi^E})
and (\ref{eq:phi^P}) due to the conservation laws of energy and
momentum. Therefore, for the processes represented in
Fig.~\ref{fig:basic-processes}(c), in addition to four zero modes
(\ref{eq:zero_modes}) and (\ref{eq:phi^J}), Eqs.~(\ref{eq:phi^E}) and
(\ref{eq:phi^P}) define two additional zero modes.  This conclusion is
consistent with six-parameter partially equilibrated distribution
(\ref{eq:n_p_partial30}).

The remaining 18 terms of the sum in Eq.~(\ref{eq:BEdiag}) are
equivalent and contain configurations of six momenta with two near one
Fermi point, and the remaining four near the opposite one, in both the
initial and final states, see Fig.~\ref{fig:basic-processes}(b).
Using the notation of the figure where $q$ and $q'$ are negative
while $k,k_1,k',k_1'$ positive, the expression in parentheses of
Eq.~(\ref{eq:BEdiag}) becomes
\begin{eqnarray}
&&\frac{\phi_{k}^{(\mathcal \eta)}}{g_{k}} + \frac{\phi_{k_1}^{(\mathcal
    \eta)}}{g_{k_1}}+ \frac{\phi_{q}^{(\mathcal \eta)}}{g_{q}}-
\frac{\phi_{k'}^{(\mathcal \eta)}}{g_{k'}}-
\frac{\phi_{k_1'}^{(\mathcal \eta)}}{g_{k_1'}}-
\frac{\phi_{q'}^{(\mathcal \eta)}}{g_{q'}}
\nonumber\\
&&=2C_\eta(q'^2-q^2),
\\
&&\frac{\phi_{k}^{(\mathcal \sigma)}}{g_{k}} +
\frac{\phi_{k_1}^{(\mathcal \sigma)}}{g_{k_1}}+
\frac{\phi_{q}^{(\mathcal \sigma)}}{g_{q}}- \frac{\phi_{k'}^{(\mathcal
    \sigma)}}{g_{k'}}- \frac{\phi_{k_1'}^{(\mathcal
    \sigma)}}{g_{k_1'}}- \frac{\phi_{q'}^{(\mathcal \sigma)}}{g_{q'}}
\nonumber\\
&&=4C_\sigma p_F(q-q'),
\end{eqnarray}
where we accounted for the conservation laws of momentum and energy. Since 
$\sum_p (\phi_p^{(\eta)})^2=
\sum_p (\phi_p^{(\sigma)})^2=L$,
where $L$ is the system size, we obtain
\begin{gather}
\label{eq:tau_eta}
\frac{1}{\tau_{\eta}}=\frac{C_\eta^2}{L}\sum_{\substack{k,k_1,k',k_1'>0\\q,q'<0}} W_{k,k_1,q}^{k',k_1',q'}(q'^2-q^2)^2,\\
\label{eq:tau_sigma}
\frac{1}{\tau_{\sigma}}=\frac{4p_F^2C_\sigma^2}{L}\sum_{\substack{k,k_1,k',k_1'>0\\q,q'<0}} W_{k,k_1,q}^{k',k_1',q'}(q'-q)^2.
\end{gather}
At low temperature, the momenta in the summation in
Eqs.~(\ref{eq:tau_eta}) and (\ref{eq:tau_sigma}) are confined near the
corresponding Fermi points. We can thus linearize
$(q'^2-q^2)^2\simeq4p_F^2(q'-q)^2$. Since $C_\eta=C_\sigma$, the two
relaxation rates (\ref{eq:tau_eta}) and (\ref{eq:tau_sigma}) are equal
at the leading order in small $T/\mu$.

\subsubsection{Evaluation of $\tau_\eta^{-1}$}

To find $\tau_\eta^{-1}$ we need an expression for the three-particle
scattering matrix element $\mathcal{A}$ that enters
Eq.~(\ref{eq:tau_eta}) via $W$ [see Eq.~(\ref{eq:W})].  The matrix
element $\mathcal{A}$ was calculated in
Ref.~\cite{ristivojevic_relaxation_2013} for an arbitrary
configuration of momenta.  In the special case of momenta that
corresponds to the process shown in Fig.~\ref{fig:basic-processes}(b),
the result of Ref.~\cite{ristivojevic_relaxation_2013} for Coulomb
interaction takes the form
\begin{align}\label{ACfinal21}
\mathcal{A}_{k,k_1,q}^{k',k_1',q'}={}&\frac{12e^4}{L^2} \frac{mw^2}{\hbar^2}\ln\left(\frac{\hbar}{p_F w}\right) \ln\biggl|\frac{k_1-k+k'-k_1'}{k-k_1+k'-k_1'}\biggr|\notag\\
&\times \delta_{k+k_1+q,k'+k_1'+q'}.
\end{align}
Similarly, for dipole-dipole interaction we have
\begin{align}
\label{eq:A-dipole-dipole}
&\mathcal{A}_{k,k_1,q}^{k',k_1',q'}=-\frac{5\Upsilon^2 m}{L^2\hbar^4}  \ln\left(\frac{\hbar}{p_F w}\right)(k_1-k)(k_1'-k')\notag\\
&\times \ln\left(\frac{p_F}{\sqrt{(k_1-k)^2+(k_1'-k')^2}}\right) \delta_{k+k_1+q,k'+k_1'+q'}.
\end{align}

We now proceed to the evaluation of the expression
(\ref{eq:tau_eta}).  Using the conservation laws of momentum and energy
we find
\begin{align}\label{eq:q-q'}
q'-q=\frac{(k'-k)(k'-k_1)}{k_1'-q}\simeq \frac{(k'-k)(k'-k_1)}{2p_F},
\end{align}
because for the process shown in Fig.~\ref{fig:basic-processes}(b),
$k_1'-q\simeq2p_F$. Therefore, we substitute
\begin{align}\label{eq:q-q'substitution}
(q'^2-q^2)^2\simeq (k'-k)^2(k'-k_1)^2
\end{align}
in Eq.~(\ref{eq:tau_eta}).  For thermally excited quasiparticles,
$|k'-k|\sim|k'-k_1|\sim T/v_F$.  On the other hand, from
Eq.~(\ref{eq:q-q'}) we find $|q'-q|\sim T^2/\mu v_F$. We therefore
find $q'=q+O(T^2/\mu v_F)$. After the substitution
(\ref{eq:q-q'substitution}), the remainder of the expression
(\ref{eq:tau_eta}) does not depend on the difference $q'-q$ apart from
the delta functions contained in $W$. We therefore approximate
\begin{align}
\label{eq:conservationlaws}
&\delta_{k+k_1+q,k'+k_1'+q'} \delta(\varepsilon_k+\varepsilon_{k_1}+\varepsilon_{q}
-\varepsilon_{k'}-\varepsilon_{k_1'}-\varepsilon_{q'}) \notag\\
&\quad\simeq\frac{1}{2v_F}\delta_{k+k_1,k'+k_1'} \delta(q-q').
\end{align} 
The integration over $q$ and $q'$ is now straightforward, resulting in
$\int dq dq' \delta(q-q') g_q g_{q'}=T/v_F$.  Here we linearized the
spectrum at low temperature, such that
\begin{align}
g_q=\frac{1}{2\cosh\frac{v_F (q+p_F)}{2T}}.
\end{align}
The remaining four integrations involve one delta function. For
Coulomb interaction (\ref{eq:Coulomb}) we were able to perform
analytically one more integration and found the relaxation rate
\begin{align}
\label{eq:rate_Coulomb}
\frac{1}{\tau_{\eta}}=c_1\frac{w^4}{a_{B}^4}\ln^2\left(\frac{\hbar }{p_F w}\right)\frac{ T^3}{\hbar \mu^2}.
\end{align}
Here $a_{B}=\hbar^2/me^2$ is the Bohr radius and
\begin{align}
c_1=-\frac{405}{4\pi^7}\int_{-\infty}^{+\infty}  \frac{u^2 w^2(u+w)\ln^2\left|\frac{w}{u}\right|du dw}{\sinh u\sinh w\sinh(u+w)}\approx 0.2306.
\end{align}
For dipole-dipole interaction (\ref{eq:dipole-dipole}) we have been
able to evaluate the numerical prefactor in the relaxation rate
analytically,
\begin{align}
\label{eq:rate_dipole-dipole}
\frac{1}{\tau_\eta}=\frac{225\pi^3}{616} \left[\frac{p_F^2 m^2\Upsilon^2}{\hbar^6} \ln\left(\frac{\hbar}{p_Fw}\right) \ln\left(\frac{\mu}{T}\right)\right]^2 \frac{ T^7}{\hbar \mu^6}.
\end{align}
Here the momentum-dependent logarithm originating from the scattering
matrix element (\ref{eq:A-dipole-dipole}) is replaced by
$\ln\left({\mu}/{T}\right)$, in accordance with the logarithmic
accuracy adopted earlier.

\subsubsection{Evaluation of $\kappa_{\rm ex}$}

As a result of separation of time scales $\tau_{\rm c}\ll \tau_{\rm b}$, at scales longer than $\tau_{\rm c}$ there exist only two relaxation modes of the distribution function given by Eqs.~(\ref{eq:phi^E}) and (\ref{eq:phi^P}). Up to a normalization constant, the former one, $\phi_{p}^{(\eta)}$ coincides with $g_p\nu_p$ of  Eq.~(\ref{eq:nu_p_approx}). Therefore the eigenmode (\ref{eq:phi^E}) actually exhausts the sum in Eq.~(\ref{eq:kappa_ex_general}) yielding 
\begin{align}
\label{eq:kappaexexpression}
\kappa_{\rm ex}=\frac{1}{4m^4 T^2} \tau_{\eta} \langle\phi_p^{(\eta)}|g_p\nu_p\rangle^2.
\end{align}
The overlap entering the latter expression can now be easily obtained by a comparison between Eqs.~(\ref{eq:nu_p_approx}) and (\ref{eq:etaapprox}):
\begin{align}
\langle\phi_p^{(\eta)}|g_p\nu_p\rangle=\frac{3p_F}{C_\eta},
\end{align}
which gives
\begin{align}
\label{eq:kappa_ex_final}
\kappa_{\rm ex}=\frac{\pi^3}{5}\frac{T^3v_F\tau_{\eta}}{\hbar \mu^2}.
\end{align}
This is our final expression for the thermal conductivity of the gas
of elementary excitations.  It applies to systems with long-range
two-body interactions.  For the Coulomb and dipole-dipole
interactions, the relaxation time $\tau_{\eta}$ is given by
Eqs.~(\ref{eq:rate_Coulomb}) and (\ref{eq:rate_dipole-dipole}).

We stress that the simplification (\ref{eq:kappa_ex_final})
corresponding to just one eigenmode of the linearized collision
integral contributing to the general expression
(\ref{eq:kappa_ex_general}) holds only for the long-range
interactions.  A different scenario occurs in systems where the
interaction potential decays rapidly with the distance.  In this case
the evaluation of $\kappa_{\rm ex}$ requires more involved study of
the relaxation modes of the collision integral, which we turn to next.

\subsection{Short-range interactions}
\label{sec:short-range}

\subsubsection{Scattering matrix element}
\label{sec:scattering-matrix_element}

In the case of short-range interactions, the rates associated with the
processes of Figs.~\ref{fig:basic-processes}(b) and (c) scale with the
temperature as $\tau_{\rm b}^{-1}\propto T^7$
\cite{imambekov_one-dimensional_2012, arzamasovs_kinetics_2014,
  protopopov_relaxation_2014} and $\tau_{\rm c}^{-1}\propto T^{14}$
\cite{protopopov_relaxation_2014}, respectively.  Thus, unlike the
cases of Coulomb and dipole-dipole interactions, at low temperatures
$\tau_{\rm b}\ll \tau_{\rm c}$.  As a result, only the processes of
Fig.~\ref{fig:basic-processes}(b) need to be taken into consideration
\cite{onefootnote}.

At low temperature $T\ll\mu$, when all the states are close to the
respective Fermi points, the scattering matrix element takes the form
\begin{equation}
  \label{eq:cal-A}
 \mathcal{A}_{k,k_1,q}^{k',k_1',q'}=\frac{\Lambda}{L^2}(k-k_1)(k'-k_1')
           \delta_{k+k_1+q,k'+k_1'+q'},
\end{equation}
where $L$ is the system size.
This expression was obtained in Ref.~\cite{matveev_decay_2013} for a
spinless quantum liquid with arbitrarily strong interactions.  It is
consistent with the matrix element used in
Ref.~\cite{khodas_fermi-luttinger_2007} provided that
\begin{equation}
  \label{eq:Lambda_weak}
  \Lambda=\frac{3V(0)V''(0)}{8mv_F^2},
\end{equation}
where $V(p)$ is the Fourier transform (\ref{eq:Fourier}) of the
interaction potential.  We note that the calculation of
Ref.~\cite{khodas_fermi-luttinger_2007} assumes that $V(p)$ falls off
rapidly away from the peak at $p=0$, such that $V(2p_F)$ is negligible
compared with $V(0)$.

In Ref.~\cite{matveev_decay_2013} the parameter $\Lambda$ was
expressed in terms of the properties of the quasiparticle spectrum of
the spinless quantum liquid.  In Appendix \ref{sec:Lambda} we apply
that prescription to the weakly-interacting spinless Fermi gas and
find
\begin{eqnarray}
  \label{eq:Lambda_Result}
  \Lambda&=&m
       \bigg(
          \frac{3 (V(0)-V(2p_F)) V''(0)}{8 p_F^2}
          +\frac{1}{2} V''(0) V''(2 p_F)
\nonumber\\
        &&
          +\frac{3V''(0) V'(2 p_F)}{4 p_F}
          -\frac{V'(2 p_F) V''(2p_F)}{4 p_F}
 \nonumber\\
        &&
          -\frac{(V'(2p_F))^2}{4 p_F^2}
         -\frac{(V(0)-V(2p_F)) V''(2 p_F)}{8 p_F^2}
\nonumber\\
        &&-\frac{(V(0)-V(2 p_F)) V'''(2p_F)}{12 p_F}
      \bigg).
\end{eqnarray}
This expression recovers Eq.~(\ref{eq:Lambda_weak}) at $V(2p_F)=0$.

\subsubsection{Linearized collision integral in the low-temperature limit}
\label{sec:linearized_collision_int_lowT}

Collision processes shown in Fig.~\ref{fig:basic-processes}(b) change
the occupation number of the state $k$ on the right-moving branch with
the rate
\begin{widetext}
\begin{eqnarray}
  \label{eq:dotn_k}
  \dot n_k&=&-\frac{2\pi}{\hbar}\,
            \frac{1}{2}\sum_{\substack{k_1,k',k_1'\\q,q'}}
            \left|\mathcal{A}_{k,k_1,q}^{k',k_1',q'}\right|^2
            \delta(\varepsilon_k+\varepsilon_{k_1}+\varepsilon_q
                   -\varepsilon_{k'}-\varepsilon_{k_1'}-\varepsilon_{q'})
\nonumber\\
           &&\qquad\qquad\quad\times \left[n_kn_{k_1}n_q(1-n_{k'})(1-n_{k_1'})(1-n_{q'})
             -(1-n_{k})(1-n_{k_1})(1-n_{q})n_{k'}n_{k_1'}n_{q'}\right].
\end{eqnarray}
Here we assume that the sums over $k'$, $k_1$, and $k_1'$ are limited
to the right-moving branch, while those over $q$ and $q'$ are limited
to the left-moving one, Fig.~\ref{fig:basic-processes}(b).  The factor
$1/2$ compensates for the double counting in the sum due to the
permutation of $k'$ and $k_1'$.  Note, that there is an additional
contribution to $\dot n_k$ due to the processes involving one
right-moving and two left-moving particles.  We will see below that at
low temperatures this contribution is negligible.

We now focus on systems close the thermal equilibrium by substituting
Eqs.~(\ref{eq:distribution_with_correction}) and (\ref{eq:x_p_definition}) and
linearizing the collision integral in small $\phi_p$.  This yields
\begin{equation}
  \label{eq:dotx_k}
  \dot \phi_k=-\frac{\pi}{\hbar}
            \sum_{\substack{k_1,k',k_1'\\q,q'}}
            \left|\mathcal{A}_{k,k_1,q}^{k',k_1',q'}\right|^2
            \delta(\varepsilon_k+\varepsilon_{k_1}+\varepsilon_q
                   -\varepsilon_{k'}-\varepsilon_{k_1'}-\varepsilon_{q'})
            g_{k_1}g_{q}g_{k'}g_{k_1'}g_{q'}
           \left(
              \frac{\phi_k}{g_k}+\frac{\phi_{k_1}}{g_{k_1}}+\frac{\phi_q}{g_q}
              -\frac{\phi_{k'}}{g_{k'}}-\frac{\phi_{k_1'}}{g_{k_1'}}
               -\frac{\phi_{q'}}{g_{q'}}
           \right).
\end{equation}
\end{widetext}

At low temperature the typical values of momentum of thermally excited
quasiparticles measured from the nearest Fermi point are of the order
of $T/v_F$.  On the other hand, as we saw in Sec.~\ref{sec:Coulomb},
for the processes of Fig.~\ref{fig:basic-processes}(b) the difference
of the momenta on the left branch is $|q-q'|\sim T^2/\mu v_F\ll
T/v_F$.  Thus, when solving Eq.~(\ref{eq:dotx_k}) to leading order in
$T/\mu\ll1$, after substituting the expression (\ref{eq:cal-A}) for
the matrix element one can apply the approximation
(\ref{eq:conservationlaws}).  This yields
\begin{eqnarray}
  \label{eq:dotx_k_simplified}
  \dot \phi_k &=&-\frac{\Lambda^2T}{32\pi^3\hbar^5 v_F^2}
              \int dk_1 dk' dk_1'
            (k-k_1)^2(k'-k_1')^2
\nonumber\\ &&\times
           \delta(k+k_1-k'-k_1')
            g_{k_1}g_{k'}g_{k_1'}
\nonumber\\ &&\times
           \left(
              \frac{\phi_k}{g_k}+\frac{\phi_{k_1}}{g_{k_1}}
              -\frac{\phi_{k'}}{g_{k'}}-\frac{\phi_{k_1'}}{g_{k_1'}}
           \right).
\end{eqnarray}
We note that because Eq.~(\ref{eq:dotx_k_simplified}) holds only in
the low-temperature limit, the same approximation must be used in
Eq.~(\ref{eq:g_p}), resulting in
\begin{equation}
  \label{eq:g_k}
  g_k=\frac{1}{2\cosh\frac{v_F(k-p_F)}{2T}}.
\end{equation}
Equation (\ref{eq:dotx_k_simplified}) shows that to leading order in
$T/\mu$ the processes involving two right-moving and one left-moving
particles [Fig.~\ref{fig:basic-processes}(b)] affect only the
distribution function on the right-moving branch.  This justifies our
earlier approximation that neglected the contribution to $\dot n_k$
from the processes involving one right-moving and two left-moving
particles.

\subsubsection{Dimensionless form of the collision integral}
\label{sec:Dimensionless}

Let us now bring Eq.~(\ref{eq:dotx_k_simplified}) to a dimensionless
form by introducing dimensionless momentum $\xi$ and time $\theta$,
\begin{equation}
  \label{eq:xi}
  \xi=\frac{v_F}{2\pi T}(k-p_F),
\quad
  \theta=\frac{2\pi^3}{\tau_\mathrm{ex}} t,
\end{equation}
where the relaxation time $\tau_{\rm ex}$ is defined by
\begin{equation}
  \label{eq:t_0}
  \frac{1}{\tau_\mathrm{ex}}=\frac{\Lambda^2T^7}{\hbar^{5} v_F^8}
\end{equation}
and has the expected power-law scaling $\tau_{\rm ex}^{-1}\propto T^7$
\cite{imambekov_one-dimensional_2012, arzamasovs_kinetics_2014,
  protopopov_relaxation_2014}.  In these units,
Eq.~(\ref{eq:dotx_k_simplified}) takes the form
\begin{equation}
  \label{eq:dotX}
  \frac{\partial \Phi(\xi)}{\partial\theta}
  =-\widehat M \Phi(\xi),
\end{equation}
where
\begin{widetext}
\begin{equation}
  \label{eq:K}
  \widehat M \Phi(\xi)
  =\int d\xi_1 d\xi' d\xi_1'
            (\xi-\xi_1)^2(\xi'-\xi_1')^2
           \delta(\xi+\xi_1-\xi'-\xi_1')
            G(\xi_1)G(\xi')G(\xi_1')
           \left(
              \frac{\Phi(\xi)}{G(\xi)}+\frac{\Phi(\xi_1)}{G(\xi_1)}
              -\frac{\Phi(\xi')}{G(\xi')}-\frac{\Phi(\xi_1')}{G(\xi_1')}
           \right).
\end{equation}
\end{widetext}
Here $\Phi(\xi)=\phi_k$ and $G(\xi)=g_k$, i.e.,
\begin{equation}
  \label{eq:G}
  G(\xi)=\frac{1}{2\cosh(\pi\xi)}.
\end{equation}
After some algebra the integral operator (\ref{eq:K}) can be rewritten
in the form
\begin{equation}
  \label{eq:K_alternative}
  \widehat M \Phi(\xi) = A(\xi) \Phi(\xi) 
           + \int d\xi'[B_1(\xi,\xi')+B_2(\xi,\xi')] \Phi(\xi'),
\end{equation}
where
\begin{eqnarray}
  \label{eq:A}
  A(\xi)&=&\frac{(1+4\xi^2)(9+4\xi^2)(5+44\xi^2)}{5760},
\\[1ex]
  \label{eq:B1}
  B_1(\xi,\xi')&=&\frac{1}{6}(\xi-\xi')^2
             \frac{(\xi+\xi')[1+(\xi+\xi')^2]}{\sinh(\pi(\xi+\xi'))},
\\[1ex]
  \label{eq:B2}
  B_2(\xi,\xi')&=&-\frac{\xi-\xi'}{240\sinh(\pi(\xi-\xi'))}
\nonumber\\ &&\times
           \big(7+120\xi\xi'+128\xi^4-752\xi^3\xi'
\nonumber\\
           &&+1488\xi^2{\xi'}^2-752\xi{\xi'}^3
           +128{\xi'}^4\big).
\end{eqnarray}
As expected, the kernel of the integral operator
(\ref{eq:K_alternative}) is symmetric with respect to permutation
$\xi\leftrightarrow\xi'$.  This property of the operator $\widehat
M$ ensures that the eigenvalue problem
\begin{equation}
  \label{eq:eigenvalue_problem_dimensionless}
  \widehat M \Phi_l(\xi)=\lambda_l \Phi_l(\xi)
\end{equation}
has an orthonormal set of solutions $\Phi_l(\xi)$ with real
eigenvalues $\lambda_l$. 

The eigenvalue problem (\ref{eq:eigenvalue_problem_dimensionless}) was
derived from Eq.~(\ref{eq:dotx_k_simplified}), which describes time
evolution of the distribution function of fermions near the right
Fermi point.  Particles near the left Fermi point can be treated in
the same way.  Therefore, each eigenfunction $\Phi_l(\xi)$ gives a
solution $\phi_p^{(l)}$ of the full eigenvalue problem
(\ref{eq:eigenvalue_problem}) that is confined to either right- or
left-moving part of the quasiparticle spectrum.  The corresponding
eigenvalue is
\begin{equation}
  \label{eq:decay_rates}
  \frac{1}{\tau_l}=\frac{2\pi^3}{\tau_\mathrm{ex}}\lambda_l.
\end{equation}
Alternatively, one can symmetrize and antisymmetrize the
eigenfunctions, resulting in two sets of solutions $\phi_p^{(l)}$ that
are either even or odd in $p$ with the same eigenvalues
(\ref{eq:decay_rates}).  Evaluation of the transport coefficient
$\kappa_{\rm ex}$ given by Eqs.~(\ref{eq:kappa_ex_general}) and
(\ref{eq:nu_p_approx}) requires odd solutions, which take the form
\begin{equation}
  \label{eq:phi_vs_Psi}
  \phi_p^{(l)}=\left(\frac{\hbar v_F}{2T}\right)^{1/2}
              \Phi_l\left(\frac{v_F(|p|-p_F)}{2\pi T}\right)\mathrm{sgn\,}p,
\end{equation}
where the prefactor assumes that the eigenfunctions $\Phi_l(\xi)$ are
normalized according to
\begin{equation}
  \label{eq:Psi_normalization}
  \int_{-\infty}^{+\infty} \Phi_l^2(\xi)\,d\xi=1.
\end{equation}

Using Eq.~(\ref{eq:phi_vs_Psi}), we can evaluate the matrix element in
our expression (\ref{eq:kappa_ex_general}) for $\kappa_\mathrm{ex}$,
\begin{equation}
  \label{eq:matrix_element-reexpressed}
  \langle\phi_p^{(l)}|g_p\nu_p\rangle
  =\frac{6\sqrt{2}\pi^2mT^{5/2}}{\hbar^{1/2}v_F^{3/2}}
   \int_{-\infty}^\infty 
   \frac{\xi^2-\frac{1}{12}}{\cosh(\pi\xi)}\Phi_l(\xi)d\xi.
\end{equation}
Substituting this matrix element into Eq.~(\ref{eq:kappa_ex_general}),
we obtain
\begin{equation}
  \label{eq:kappa_ex_transformed}
  \kappa_{\rm ex}=\frac{9\pi c_2}{4}
                \, \frac{T^3v_F\tau_\mathrm{ex}}{\hbar \mu^2},
\end{equation}
where $\tau_{\rm ex}$ is given by Eq.~(\ref{eq:t_0}) and the numerical
coefficient $c_2$ is defined as
\begin{equation}
  \label{eq:c}
  c_2=\sideset{}{'}\sum_{l} \frac{1}{\lambda_l} 
         \bigg[\int_{-\infty}^\infty 
   \frac{\xi^2-\frac{1}{12}}{\cosh(\pi\xi)}\Phi_l(\xi)d\xi\bigg]^2.
\end{equation}
Summation in Eq.~(\ref{eq:c}) excludes the zero modes for which
$\lambda_l=0$.

Because the dimensionless eigenvalue problem
(\ref{eq:eigenvalue_problem_dimensionless}) describes relaxation of
the right-moving particles, which at $T/\mu\to0$ is decoupled from the
relaxation of the left movers, operator $\widehat M$ has only two zero
modes:
\begin{eqnarray}
  \label{eq:X0}
  \Phi_0(\xi)&=&\sqrt{2\pi}\,G(\xi),
\\
  \label{eq:X1}
  \Phi_1(\xi)&=&\sqrt{24\pi}\,\xi\, G(\xi).
\end{eqnarray}
The modes $\Phi_0$ and $\Phi_1$ correspond to the conservation of the
particle number and momentum, respectively.  Their forms are easily
verified analytically using Eq.~(\ref{eq:K}).  The integral operator
(\ref{eq:K_alternative}) can be diagonalized numerically.  Excluding
the zero modes (\ref{eq:X0}) and (\ref{eq:X1}) from the sum
(\ref{eq:c}), we obtained $c_2\approx0.41088$.

\section{Discussion of the results}
\label{sec:discussion}

In this paper we have developed a microscopic theory of the thermal
transport coefficients of one-dimensional Fermi gas at low temperature
$T\ll \mu$.  A special feature of one-dimensional quantum systems is
that in addition to the usual thermal conductivity of the system
$\kappa$, one can introduce thermal conductivity of the gas of
elementary excitations $\kappa_{\rm ex}$.  This is a consequence of
separation of scales of the rates of various processes responsible for
the relaxation of the system to equilibrium.  Specifically, the rates
of the processes of Fig.~\ref{fig:basic-processes}(a), which are
responsible for the equilibration of the chemical potentials of the
left- and right-moving particles, are exponentially small,
$\tau^{-1}\propto e^{-\mu/T}$.  On the other hand, the remaining
scattering processes illustrated in Fig.~\ref{fig:basic-processes}(b)
and (c) occur at rates $\tau_{\rm ex}^{-1}$ that scale as a power law
of $T/\mu$ and are therefore much faster.

In general, transport coefficients are proportional to the relevant
relaxation times.  As a result, the thermal conductivity
$\kappa\propto\tau$ is exponentially large.  Our microscopic theory
gives the result (\ref{eq:kappa_low_T}), which is consistent with the
phenomenological expression (\ref{eq:kappa_Luttinger}) obtained
within the Luttinger liquid theory.  These results relate $\kappa$ to
the relaxation time $\tau$, which requires special evaluation.  A
phenomenological theory \cite{matveev_equilibration_2012,
  matveev_scattering_2012} expresses $\tau$ in terms of the properties
of the excitation spectrum of the quantum liquid.  For the system of
weakly interacting fermions further progress can be made.  In
Appendix~\ref{sec:tau_prefactor} we obtained an expression for $\tau$
in terms of the interaction potential.  For example, in the case of
spinless fermions with dipole-dipole interaction
(\ref{eq:dipole-dipole}) we found
\begin{equation}
  \label{eq:tau_dipole-dipole}
  \frac{1}{\tau}=\frac{36(\ln4-1)^2}{5\pi^{3/2}}
                 \frac{\Upsilon^4p_F^9}{\hbar^{13}v_F^3}
                 \ln^2\left(\frac{\hbar}{p_F w}\right)
                 \left(\frac{T}{\mu}\right)^{3/2}
                 e^{-\mu/T}.
\end{equation}
A more general expression given by Eqs.~(\ref{eq:tau_result}) and
(\ref{eq:Z}) applies to any interaction $U(x)$ that falls off at large
distances faster than $1/x$.

In one-dimensional systems the transport coefficient $\kappa$
describes the thermal conductivity only at low frequencies
$\omega\ll\tau^{-1}$.  In particular, it gives the dominant
contribution to the attenuation of sound at low frequencies
\cite{matveev_propagation_2018}.  On the other hand, at $\omega\gg
\tau^{-1}$ the exponentially slow relaxation processes can be
neglected, which leads to a very different behavior of the system.
Instead of the conventional sound, the system now supports two sound
modes \cite{matveev_second_2017, matveev_hybrid_2018}, whose
attenuation is no longer exponentially strong.  Instead, the thermal
transport coefficient that enters the expression for sound attenuation
is $\kappa_{\rm ex}$ \cite{matveev_propagation_2018}, which can be
thought of as the thermal conductivity at frequencies $\omega \gg
\tau^{-1}$.

Our microscopic theory relates $\kappa_{\rm ex}$ to the relaxation
modes of the system, see Eqs.~(\ref{eq:kappa_ex_general}) and
(\ref{eq:nu_p_approx}).  This enables one to find the order of
magnitude estimate (\ref{eq:kappa_ex_estimate}), which expresses
$\kappa_{\rm ex}$ in terms of the relaxation time $\tau_{\rm ex}$.  To
obtain a full microscopic expression for $\kappa_{\rm ex}$, a more
detailed treatment of the relaxation processes is required.  We
performed such a treatment for two kinds of interactions between
fermions.  For Coulomb and dipole-dipole interactions, the long range
of the interaction potential results in processes of the type shown in
Fig.~\ref{fig:basic-processes}(c) having a higher rate than those of
Fig.~\ref{fig:basic-processes}(b).  The relaxation rate
$\tau_{\eta}^{-1}$ that controls the thermal conductivity of the gas
of excitations is due to the latter type of processes and is given by
Eqs.~(\ref{eq:rate_Coulomb}) and (\ref{eq:rate_dipole-dipole}).  Our
microscopic result for $\kappa_{\rm ex}$ is given by
Eq.~(\ref{eq:kappa_ex_final}).  It is consistent with the earlier
estimate (\ref{eq:kappa_ex_estimate}) provided that the rate
$\tau_{\rm ex}^{-1}$ of relaxation of the gas of excitations is
identified with $\tau_{\eta}^{-1}$.  On the other hand, in the case of
short-range interactions, the processes shown in
Fig.~\ref{fig:basic-processes}(c) are negligible.  The resulting
$\kappa_{\rm ex}$ is expressed in terms of the interaction potential
using Eqs.~(\ref{eq:Lambda_Result}), (\ref{eq:t_0}) and
(\ref{eq:kappa_ex_transformed}).

In our treatment we considered the model of spinless fermions.  Spins
can be added to the Boltzmann equation treatment of thermal transport
in a straightforward way, but they may strongly affect the relaxation
processes.  The processes of Fig.~\ref{fig:basic-processes}(a) remain
exponentially suppressed in the presence of spins, and we again expect
$\kappa\propto\tau\propto e^{\mu/T}$.  The rate of processes shown in
Fig.~\ref{fig:basic-processes}(b) was studied in
Ref.~\cite{karzig_energy_2010}, where $\tau_{\rm ex}^{-1}\propto T$
was obtained, while the processes of Fig.~\ref{fig:basic-processes}(c)
have not been explored.  In the case of spinless fermions the
relaxation rate relevant for the evaluation of $\kappa_{\rm ex}$ was
that of the processes shown in Fig.~\ref{fig:basic-processes}(b), and
this should hold for fermions with spin.  Thus, using the estimate
(\ref{eq:kappa_ex_estimate}) we expect that for weakly interacting
spin-$\frac12$ fermions $\kappa_{\rm ex}\propto T^2$.  We leave more
careful microscopic treatment of this problem for future study.

Finally, we note that our treatment of the thermal conductivity
assumes that at every point in space the Fermi gas is near local
thermal equilibrium.  This assumption is violated if one takes into
account the sound modes in the system, which are weakly damped at low
frequencies \cite{andreev_two-liquid_1971}.  This effect is
particularly strong in one dimension, where it leads to power-law
scaling of sound absorption \cite{andreev_hydrodynamics_1980}.
Violation of the assumption of local equilibrium gives rise to
power-law scaling of thermal conductance of one-dimensional classical
systems with the system size, see, e.g.,
Ref.~\cite{lepri_thermal_2003}.  This effect should also manifest
itself in the low-frequency behavior of thermal conductivity of
one-dimensional Fermi gas at low temperatures.  On the other hand,
because the local contribution (\ref{eq:kappa_Luttinger}) to the
thermal conductivity is exponentially amplified by the long relaxation
time $\tau$, one should expect the non-local effects to become
significant only at frequencies that are exponentially smaller than
$\tau^{-1}$.  Frequency dependence of thermal conductivity of
one-dimensional electronic fluid was recently studied by R. Samanta
\emph{et al.}~\cite{samantha_thermal_2019}.  In addition to the
crossover from the result (\ref{eq:kappa_Luttinger}) to a power-law
scaling at $\omega\to0$, they obtained several additional parametric
frequency regions, including one in which $\kappa(\omega)$ is
consistent with our result for $\kappa_{\rm ex}$.

\begin{acknowledgments}

  The authors are grateful to A. V. Andreev for stimulating
  discussions.  Work at Argonne National Laboratory was supported by
  the U.S. Department of Energy, Office of Science, Materials Sciences
  and Engineering Division.  Work at Laboratoire de Physique
  Th\'{e}orique was supported in part by the EUR grant NanoX
  ANR-17-EURE-0009 in the framework of the ``Programme des
  Investissements d’Avenir.''

\end{acknowledgments}

\appendix

\section{Fourier transform of the Coulomb interaction potential}
\label{sec:Fourier}

Let us consider two three-dimensional particles with charge $e$
confined to a one-dimensional channel.  The position of each particle
will be described by its coordinate along the channel and the vector
$\bm\rho$ in the transverse direction.  Coulomb interaction between
the two particles at the distance $x$ along the channel is given by
\begin{equation}
  \label{eq:Coulomb_general}
  U(x)=e^2\!\int\!\!\!\int\! d\bm\rho_1 d\bm\rho_2
       \frac{|\Psi(\bm\rho_1)|^2|\Psi(\bm\rho_2)|^2}
            {\sqrt{x^2+|\bm\rho_1-\bm\rho_2|^2}},
\end{equation}
where $\Psi(\bm\rho)$ is the normalized wave function of the
transverse motion of a particle in the channel.  Performing the
Fourier transform (\ref{eq:Fourier}), we find
\begin{equation}
  \label{eq:Fourier_exact}
  V(p)=2e^2\!\int\!\!\!\int\! d\bm\rho_1 d\bm\rho_2
       |\Psi(\bm\rho_1)|^2|\Psi(\bm\rho_2)|^2
       K_0\Big(\frac{p}{\hbar}|\bm\rho_1-\bm\rho_2|\Big),
\end{equation}
where $K_0(z)$ is the Macdonald function.  Its asymptotic behavior at
small $z$ is given by
\begin{equation}
  \label{eq:Macdonald_expansion}
  K_0(z)=\ln\frac{2}{z}-\bm C 
         +\frac{z^2}{4}\bigg(\ln\frac{2}{z}+1-\bm C\bigg)
         +o(z^3),
\end{equation}
where $\bm C\approx0.5772$ is the Euler's constant.  Substituting
Eq.~(\ref{eq:Macdonald_expansion}) into Eq.~(\ref{eq:Fourier_exact})
and omitting coefficients of order unity in the argument of the
logarithms, we obtain the result (\ref{eq:Coulomb}), provided that the
channel width $w$ is defined by
\begin{equation}
  \label{eq:w^2}
  w^2=\frac{1}{2}\Big(\overline{\bm\rho^2}-\overline{\bm\rho}^2\Big).
\end{equation}
Here the averaging 
\begin{equation}
  \label{eq:transverse_averaging}
  \overline{f(\bm\rho)}=\int\! d\bm\rho\,|\Psi(\bm\rho)|^2f(\bm\rho)
\end{equation}
is performed over the distribution of the particle density in the
transverse direction.

\section{Evaluation of $\Lambda$ in the limit of weak interactions}
\label{sec:Lambda}

In this Appendix we evaluate the parameter $\Lambda$ in the expression
for the scattering matrix element (\ref{eq:cal-A}) characterizing the
process shown in Fig.~\ref{fig:basic-processes}(b).  We will express
$\Lambda$ in terms of the Fourier components $V(p)$ of the interaction
potential and obtain the result (\ref{eq:Lambda_Result}).  Our
calculation is based on the approach suggested in
Ref.~\cite{matveev_decay_2013} where the matrix element
(\ref{eq:cal-A}) was used to study the decay rate of quasiparticle
excitations of a spinless quantum liquid at zero temperature.  The
prescription of Ref.~\cite{matveev_decay_2013} is
\begin{equation}
  \label{eq:Lambda_prescription}
  \Lambda=-\lim_{p\to p_F}\frac{Y_p}{(p-p_F)^2},
\end{equation}
where $Y_p$ is expressed in terms of the quasiparticle energies
$\epsilon_p$ as follows
\begin{eqnarray}
  \label{eq:Y_p}
      &&Y_p=
      \partial^2_{LR}\epsilon_p
      -\frac{1}{m_p^*}
      \frac{\partial_L\epsilon_{p}}{v+v_p}
      \frac{\partial_R\epsilon_{p}}{v-v_p}
     +\partial_L v_p
      \frac{\partial_R\epsilon_{p}}{v-v_p}
\nonumber\\
   &&\hspace{0.5em} -\partial_R v_p
      \frac{\partial_L\epsilon_{p}}{v+v_p}
      +\frac{v\partial_n K}{\sqrt K}
      \bigg(
       \frac{\partial_R\epsilon_p}{v-v_p}
       +\frac{\partial_L\epsilon_p}{v+v_p}
      \bigg).
\end{eqnarray}
Here the quasiparticle velocity is defined as a derivative of
quasiparticle energy, $v_p=\epsilon_p'$, the effective
mass $m^*_p=1/\epsilon_p''$, the speed of the low energy
excitations $v=v_{p_F}^{}$, and for Galilean invariant systems the
Luttinger liquid parameter $K=v_F/v$.  In an interacting system the
quasiparticle energies depend on the density of particles $n$ and
momentum per particle $\chi$.  The partial derivatives in
Eq.~(\ref{eq:Y_p}) are defined by
\begin{subequations}
\label{eq:derivatives}
\begin{eqnarray}
  \hspace{-2em}
   \partial_R &=& \sqrt{K}\partial_n
             +\frac{\pi\hbar}{\sqrt K}\partial_\chi,
\\
   \partial_L &=& \sqrt{K}\partial_n
             -\frac{\pi\hbar}{\sqrt K}\partial_\chi,
\\
  \partial^2_{LR} &=& K\partial_n^2-\frac{(\pi\hbar)^2}{K}\partial_\chi^2.
\end{eqnarray}
\end{subequations}
Finally, since the sign of $\Lambda$ has no physical significance, for
convenience, in Eq.~(\ref{eq:Lambda_prescription}) we changed the sign
of the expression used in Ref.~\cite{matveev_decay_2013}.

A quasiparticle can be added to the ground state of the system in two
ways.  First, one can move a fermion from a Fermi point $p_F$ to a
state with momentum $p>p_F$.  In this approach, the quasiparticle is
essentially a particle-hole pair, with the hole remaining at the Fermi
point.  Alternatively, an additional fermion with momentum $p>p_F$ can
be added to the system.  The former approach was used in
Ref.~\cite{matveev_decay_2013}, but the prescription
(\ref{eq:Lambda_prescription}) and (\ref{eq:Y_p}) applies in both
cases.  In this Appendix we will use the second approach.  In
particular, the energy of the excitation in the system of
non-interacting fermions is
\begin{equation}
  \label{eq:epsilon_p}
  \epsilon_p^{(0)}=\frac{p^2}{2m}.
\end{equation}
This energy, of course, depends on neither the density $n$ nor the
momentum per particle $\chi$ of the system, and therefore yields
$\Lambda=0$.  In a weakly interacting system, the energies of both the
ground state and the state with the additional particle with momentum
$p$ change.  In this case the quasiparticle energy should be
understood as the difference of the energies of those many-body
states.  The energy $\epsilon_p$ defined this way does depend on $n$
and $\chi$, resulting in a nonvanishing $\Lambda$.

Instead of treating quasiparticle energies as functions of $n$ and
$\chi$, it will be convenient to think of the ground state of a moving
system in terms of the positions of the left and right Fermi points,
$p_L$ and $p_R$.  The two sets of variables are related by
\begin{equation}
  \label{eq:density_and_kappa}
  n=\frac{p_R-p_L}{2\pi\hbar},
  \qquad
  \chi=\frac{p_R+p_L}{2}.
\end{equation}
Then the derivatives (\ref{eq:derivatives}) can be written as
\begin{subequations}
\label{eq:derivatives_new}
\begin{eqnarray}
  \hspace*{-5em}
   \partial_R &=& \pi\hbar\left(\frac{1}{\sqrt K}+\sqrt{K}\right)
                   \frac{\partial}{\partial_{p_R}}
\nonumber\\ &&
             +\pi\hbar\left(\frac{1}{\sqrt K}-\sqrt{K}\right)
                   \frac{\partial}{\partial_{p_L}},
\\
   \partial_L &=& -\pi\hbar\left(\frac{1}{\sqrt K}-\sqrt{K}\right)
                   \frac{\partial}{\partial_{p_R}}
\nonumber\\ &&
             -\pi\hbar\left(\frac{1}{\sqrt K}+\sqrt{K}\right)
                   \frac{\partial}{\partial_{p_L}},
\\
  \partial^2_{LR} &=& -2\pi^2\hbar^2\bigg(\frac{1}{K}+K\bigg)
                      \frac{\partial^2}{\partial_{p_R}\partial_{p_L}}
\nonumber\\ &&
             -\pi^2\hbar^2\left(\frac{1}{K}-K\right)
               \left(\frac{\partial^2}{\partial_{p_R^2}}
                      +\frac{\partial^2}{\partial_{p_L^2}}\right).
\label{eq:second_derivative_new}
\end{eqnarray}
\end{subequations}
In the following, we evaluate the quasiparticle energy
$\epsilon_p(p_L,p_R)$ up to terms quadratic in interactions,
substitute the resulting expressions into Eq.~(\ref{eq:Y_p}), take the
limits $p_L\to-p_F$ and $p_R\to p_F$, and obtain $\Lambda$ from
Eq.~(\ref{eq:Lambda_prescription}).

We start by evaluating the first order correction
$\delta\epsilon_k^{(1)}$ to the quasiparticle energy
(\ref{eq:epsilon_p}).  The interaction between fermions is given by
\begin{equation}
  \label{eq:interaction}
  \widehat V = \frac{1}{2L}\sum_{p_1,p_2,q} 
               V(q)a_{p_1+q}^\dagger a_{p_2-q}^\dagger a_{p_2} a_{p_1},
\end{equation}
where $V(q)$ is defined by Eq.~(\ref{eq:Fourier}).  The first-order
correction to the energy of a many-body state
\begin{equation}
  \label{eq:E1}
  \delta E^{(1)}=\frac{1}{2L}\sum_{p_1,p_2} 
               [V(0)-V(p_1-p_2)]n_{p_1}n_{p_2},
\end{equation}
where $n_p=\langle a_p^\dagger a_p^{}\rangle$ is the occupation number
of the state with momentum $p$.  In a moving ground state
\begin{equation}
  \label{eq:n_p_moving}
  n_p=\theta(p-p_L)-\theta(p-p_R),
\end{equation}
where $\theta(x)$ is the unit step function, and we assumed $p_R>p_L$.
The extra contribution to $\delta E^{(1)}$ from an additional fermion
at state $p$ is given by
\begin{eqnarray}
  \label{eq:varepsilon1}
  \delta\epsilon^{(1)}_p=\frac{1}{2\pi\hbar}\int_{p_L}^{p_R}dp_1[V(0)-V(p-p_1)].
\end{eqnarray}
Differentiating this expression with respect to $p$, we find the first
order correction to quasiparticle velocity $v_p^{(0)}=p/m$ in the form
\begin{eqnarray}
  \label{eq:v1}
   \delta v_p^{(1)}=\frac{V(p-p_R)-V(p-p_L)}{2\pi\hbar}.
\end{eqnarray}

\begin{figure}[t]
\includegraphics[width=.22\textwidth]{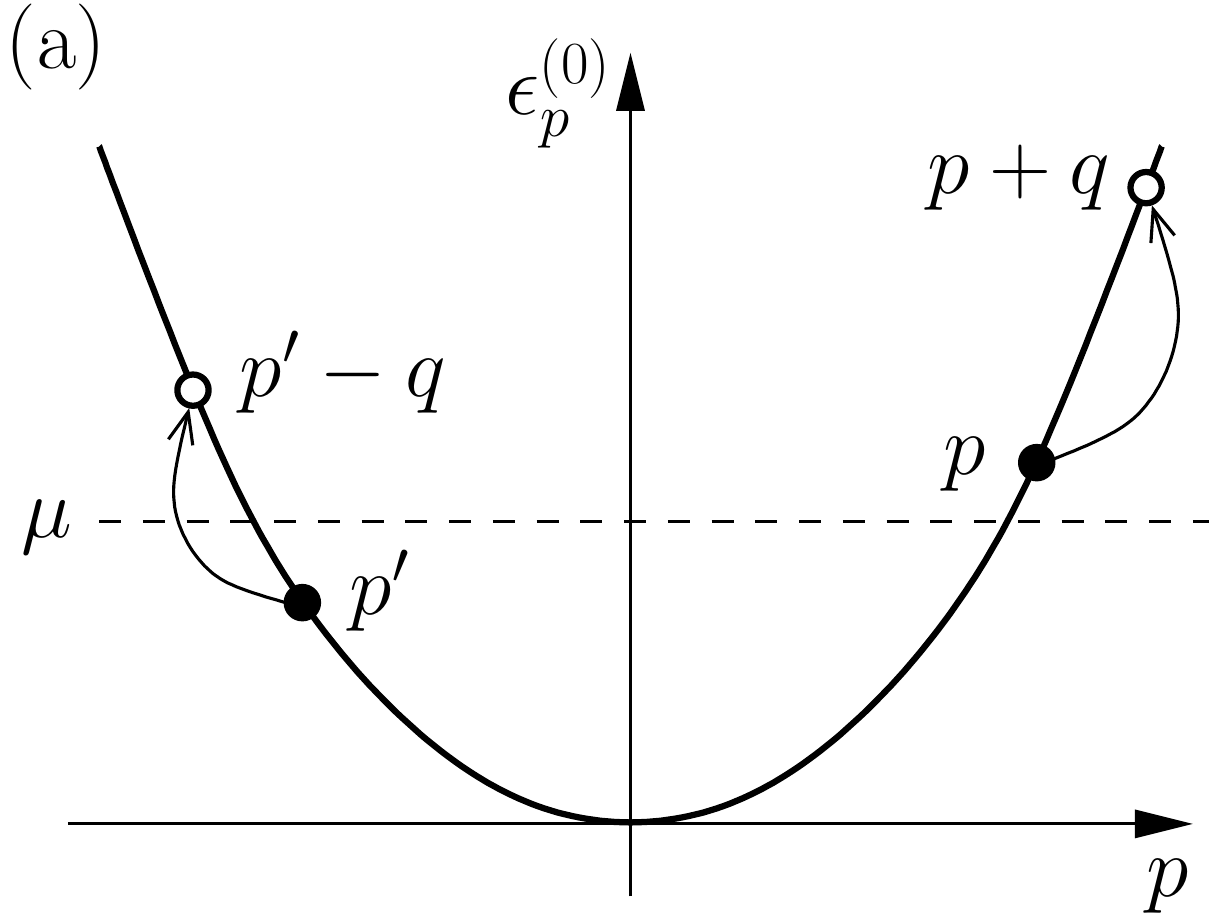}
\quad\quad
\includegraphics[width=.22\textwidth]{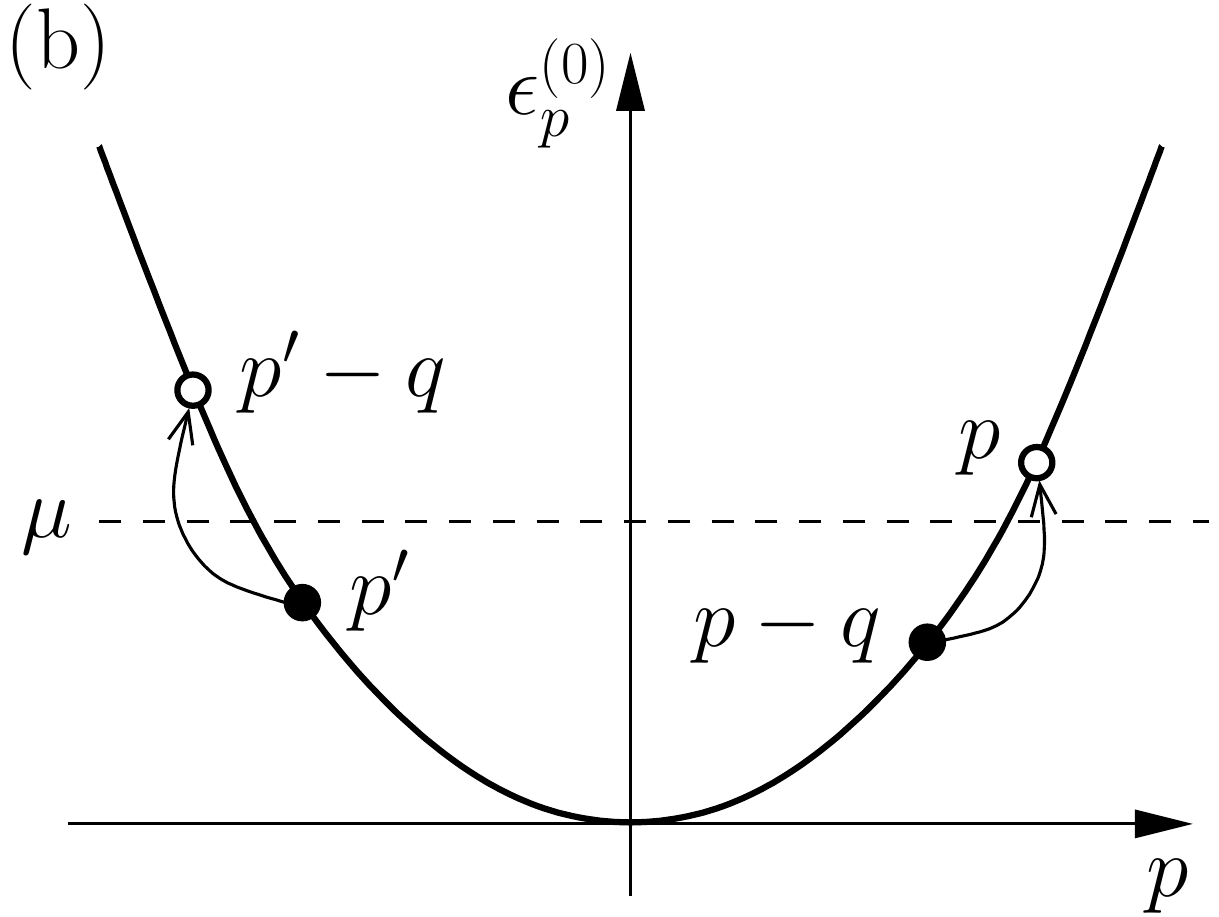}
\caption{Second-order contributions to the quasiparticle energy.
  Processes (a) and (b) correspond to the first and second terms in
  Eq.~(\ref{eq:varepsilon2}), respectively.}
\label{fig:second-order}
\end{figure}

In a stationary system, $p_R=-p_L=p_F$, the velocity at the Fermi
point is
\begin{equation}
  \label{eq:v_F1}
  v=v_F+\frac{V(0)-V(2p_F)}{2\pi\hbar}.
\end{equation}
As a result, to first order in interaction the Luttinger liquid
parameter $K=v_F/v$ given by
\begin{equation}
  \label{eq:K1}
  K=1-\frac{V(0)-V(2p_F)}{2\pi\hbar v_F}.
\end{equation}
Because we have $K=1$ in the absence of interactions, the derivative
of $K$ with respect to the particle density appears in the first order
in interactions,
\begin{equation}
  \label{eq:d_n_K}
  \partial_nK=\frac{1}{2 p_F v_F}[V(0)-V(2p_F)+2p_FV'(2p_F)].
\end{equation}
Similarly, the derivatives of $\epsilon_p$ and $v_p$ in
Eq.~(\ref{eq:Y_p}) appear only in the first order.  These leading
contributions are found by using Eq.~(\ref{eq:derivatives_new}),
(\ref{eq:varepsilon1}), and (\ref{eq:v1}),
\begin{subequations}
  \label{eq:first_order_derivatives}
\begin{eqnarray}
  \partial_R\epsilon_p^{(1)}&=&V(0)-V(p-p_R),
\\
  \partial_L\epsilon_p^{(1)}&=&V(0)-V(p-p_L),
\\
  \partial_Rv_p^{(1)}&=&-V'(p-p_R),
\\
  \partial_Lv_p^{(1)}&=&-V'(p-p_L).
\end{eqnarray}
\end{subequations}
On the other hand, the second derivative
$\partial^2_{LR}\epsilon_p^{(1)}$ appears only in the second order
in interactions
\begin{eqnarray}
  \label{eq:d2varepsilon1}
  \partial^2_{LR}\delta\epsilon_p^{(1)}&=&-\pi^2\hbar^2\left(\frac{1}{K}-K\right)
               \left(\frac{\partial^2}{\partial_{p_R^2}}
                      +\frac{\partial^2}{\partial_{p_L^2}}\right)
                  \delta\epsilon_p^{(1)}
\nonumber\\
   &\simeq&
   -\frac{V(0)-V(2p_F)}{2 v_F}[V'(p-p_R)-V'(p-p_L)].
\nonumber\\
\end{eqnarray}
Substituting Eqs.~(\ref{eq:d_n_K})--(\ref{eq:d2varepsilon1}) into
Eq.~(\ref{eq:Y_p}), we conclude that to first order in interactions
$Y_p$ and, therefore, $\Lambda$ vanish.  This is an expected outcome
because the three-particle scattering matrix element (\ref{eq:cal-A})
cannot be generated from the two-particle interaction
(\ref{eq:interaction}) in the first order.

To obtain $Y_p$ in the second order in interaction strength, in
addition to Eqs.~(\ref{eq:d_n_K})--(\ref{eq:d2varepsilon1}) we need to
find the second-order correction $\delta\epsilon_p^{(2)}$ to the
quasiparticle energy, which will contribute to the first term in the
right-hand side of Eq.~(\ref{eq:Y_p}).  Applying the standard
second-order perturbation theory, we obtain
\begin{widetext}
\begin{eqnarray}
  \label{eq:varepsilon2}
  \delta\epsilon_p^{(2)}&=&\frac{1}{(2\pi\hbar)^2}
         \int\!\!\int dp'\,dq\,
         \Bigg(
         n_{p'}(1-n_{p'-q})(1-n_{p+q})
         \frac{[V(q)-V(p+q-p')]^2}
         {\epsilon_{p}^{(0)}+\epsilon_{p'}^{(0)}
          -\epsilon_{p+q}^{(0)}-\epsilon_{p'-q}^{(0)}}
\nonumber\\
         &&\hspace{8em}
         -n_{p'}(1-n_{p'-q})n_{p-q}
         \frac{[V(q)-V(p-p')]^2}
         {\epsilon_{p-q}^{(0)}+\epsilon_{p'}^{(0)}
          -\epsilon_{p}^{(0)}-\epsilon_{p'-q}^{(0)}}
         \Bigg)
\end{eqnarray}
The first term in the integral accounts for the second-order
contribution to the many-body state due to an additional fermion in
state $p$, assuming $p>p_F$.  The second term subtracts the
contribution to the ground state energy due to processes involving
state $p$, which are not allowed in the presence of the additional
fermion.  The two contributions are illustrated in
Fig.~\ref{fig:second-order}.

The second-order correction (\ref{eq:varepsilon2}) contributes to the
first term in Eq.~(\ref{eq:Y_p}) via
$\partial^2_{LR}\epsilon_p$.  To leading order in interactions,
$\partial^2_{LR}\delta\epsilon_p^{(2)}=
-(2\pi\hbar)^2 \partial^2\epsilon_p^{(2)}/\partial_{p_R}\partial_{p_L}$,
see Eq.~(\ref{eq:second_derivative_new}).  The positions $p_L$ and
$p_R$ of the two Fermi points enter Eq.~(\ref{eq:varepsilon2}) via the
occupation numbers (\ref{eq:n_p_moving}).

Differentiation of Eq.~(\ref{eq:varepsilon2}) with respect to $p_L$
and $p_R$ yields
\begin{eqnarray}
  \label{eq:d2varepsilon2}
   \partial^2_{LR}\delta\epsilon_p^{(2)}&=&
      -\frac{[V(2p_F)-V(p+p_F)]^2}
         {\epsilon_p^{(0)}-\epsilon_{p+2p_F}^{(0)}}
      -\frac{[V(p-p_F)-V(p+p_F)]^2}
         {2\Big(\epsilon_{p_F}^{(0)}-\epsilon_{p}^{(0)}\Big)}
      +\frac{[V(2p_F)-V(p-p_F)]^2}
         {\epsilon_{p-2p_F}^{(0)}-\epsilon_{p}^{(0)}}.
\end{eqnarray}
Here the first term originates from the first term in
Eq.~(\ref{eq:varepsilon2}), whereas the second and third ones
originate from the second term in (\ref{eq:varepsilon2}).

We are now in a position to evaluate $Y_p$ in the second order in
interaction strength.  To this end we substitute Eqs.~(\ref{eq:d_n_K})
and (\ref{eq:first_order_derivatives}) for the corresponding
derivatives in the first order and add expressions
(\ref{eq:d2varepsilon1}) and (\ref{eq:d2varepsilon2}) evaluated in the
second order.  (We replace $p_L\to-p_F$ and $p_R\to p_F$.)  The
remaining parameters need not account for interactions, i.e., we
substitute $m^*_p= m$, $v= v_F$, $v_p= p/m$, and $K=1$.  The
result has the form
\begin{eqnarray}
  \label{eq:Y_p_result}
  Y_p&=&m
        \bigg[
         \frac{[V(2p_F)-V(p+p_F)]^2}
         {2p_F(p+p_F)}
          -\frac{[V(2p_F)-V(p-p_F)]^2}
         {2p_F(p-p_F)}
          +\frac{[V(p-p_F)-V(p+p_F)]^2}
         {(p+p_F)(p-p_F)}
\nonumber\\
      &&-\frac{V(0)-V(2p_F)}{2p_F}[V'(p-p_F)-V'(p+p_F)]
        +\frac{[V(0)-V(p-p_F)][V(0)-V(p+p_F)]}{(p+p_F)(p-p_F)}
\nonumber\\
      &&+\frac{V(0)-V(p-p_F)}{p-p_F}V'(p+p_F)
        +\frac{V(0)-V(p+p_F)}{p+p_F}V'(p-p_F)
\nonumber\\
      &&+\left(
          \frac{V(0)-V(2p_F)}{2p_F}+V'(2p_F)
         \right)
         \left(
          \frac{V(0)-V(p+p_F)}{p+p_F}-\frac{V(0)-V(p-p_F)}{p-p_F}
         \right)
        \bigg].
\end{eqnarray}
\end{widetext}
Substitution of the above result into
Eq.~(\ref{eq:Lambda_prescription}) yields
Eq.~(\ref{eq:Lambda_Result}).

\section{Relaxation rate $\tau^{-1}$ in a system of weakly interacting spinless fermions}
\label{sec:tau_prefactor}

A phenomenological expression for the relaxation rate of in a spinless
quantum liquid at low temperatures was obtained in
Refs.~\cite{matveev_equilibration_2012, matveev_scattering_2012}.  In
this Appendix we apply that result to find the rate $\tau^{-1}$ for
the special case of weakly interacting spinless fermions.  To this end
it is convenient to express the relaxation rate as
\cite{matveev_equilibration_2013}
\begin{equation}
  \label{eq:tau_general}
  \frac1\tau=\frac{3B}{\pi^{5/2}p_F^2}
           \left(\frac{vp_F}{T}\right)^3
           \left(\frac{p_F^2}{2 m^* T}\right)^{1/2}
           e^{-\Delta/T}.
\end{equation}
Here $\Delta$ is the maximum energy of a hole-like excitation in the
quantum liquid, $m^*$ is the effective mass of the hole at the maximum
of energy, and $v$ is the velocity of the low-energy excitations in
the system.  In a weakly interacting Fermi gas $\Delta=\mu$, $m^*=m$,
and $v=v_F$.

The quantity $B$ was expressed in
Ref.~\cite{matveev_equilibration_2012} in terms of $\Delta$, $m^*$,
and $v$ as functions of the particle density.  An alternative
expression
\begin{equation}
  \label{eq:B}
  B=\frac{4\pi}{15}\frac{Y_0^2}{\hbar^5 v^{6}}\,T^5
\end{equation}
was obtained in Refs.~\cite{matveev_scattering_2012,
  matveev_equilibration_2013}.  Here $Y_0$ is a function of the
spectrum of holes in the quantum liquid, which is analogous to the
$Y_p$ for particle-like excitations given by Eq.~(\ref{eq:Y_p}).
Because of the difference in the type of excitations, the effective
mass $m^*$ in Eq.~(\ref{eq:Y_p}) should be replaced with $-m^*$.  The
momentum $p$ in the resulting $Y_p$ should correspond to the maximum
of the energy of the hole.  If the hole is formed by moving a fermion
from a state below the Fermi level to the right Fermi point, the
maximum of energy corresponds to $p=p_F$.  Alternatively, one may
create a hole by removing a particle from the system, in which case
the maximum of energy corresponds to $p=0$.  Here we adopt the latter
approach.

Evaluation of $Y_p$ for a hole-like excitation can be performed by
retracing the steps leading from Eq.~(\ref{eq:Y_p}) to
Eq.~(\ref{eq:Y_p_result}).  We find that for the hole $Y_p$ is given
by Eq.~(\ref{eq:Y_p_result}) with the opposite sign.  Thus one
can substitute for $Y_0$ into Eq.~(\ref{eq:B}) the result
(\ref{eq:Y_p_result}) taken at $p=0$.  This yields
\begin{equation}
  \label{eq:tau_result}
  \frac1\tau=\frac{2}{5\pi^{3/2}}\,
           \frac{\mu Z^2}{\hbar^5 v_F^4}
           \left(\frac{T}{\mu}\right)^{3/2}
           e^{-\mu/T},
\end{equation}
where
\begin{eqnarray}
  \label{eq:Z}
  Z&=&[V(0)-V(2p_F)][V(p_F)-V(2p_F)]
\nonumber\\
    &&+2p_F[V(0)-V(p_F)]V'(2p_F)
\nonumber\\
    &&-p_F[V(0)-2V(p_F)+V(2p_F)]V'(p_F).
\end{eqnarray}

The derivation of the expression (\ref{eq:tau_general}) in
Refs.~\cite{matveev_equilibration_2012, matveev_scattering_2012}
assumed that the interactions between the particles fall off
sufficiently fast at the long distances for the velocity $v$ of the
elementary excitations to be well defined.  In practice this means
that the interaction potential $U(x)$ falls off at $x\to\infty$ faster
than $1/x$.  In particular, the result (\ref{eq:tau_result}) does not
apply in the case of Coulomb interactions, for which $V(0)$ in
Eq.~(\ref{eq:Z}) is ill-defined.  On the other hand, the dipole-dipole
interaction with the short-distance cutoff $w$ does have a
well-defined $V(0)$.  Substituting Eq.~(\ref{eq:dipole-dipole}) into
Eq.~(\ref{eq:Z}), we obtain
\begin{equation}
  \label{eq:Z_dipole-dipole}
  Z=-6(\ln4-1)\Upsilon^2\frac{p_F^4}{\hbar^4}
     \ln\frac{\hbar}{p_Fw}.
\end{equation}
This expression is obtained for $p_Fw/\hbar\ll1$ within logarithmic
accuracy.


\begin{thebibliography}{26}
\bibitem [{one()}]{onefootnote}%
  \BibitemOpen
  \href@noop {} {}\bibinfo {note} {Even though the processes of
  Fig.~\ref{fig:basic-processes}(b) are weaker than those of
  Fig.~\ref{fig:basic-processes}(c) for the long-range interactions, they
  played an important role in Sec.~\ref{sec:Coulomb}, because the dominant
  processes in that case did not equilibrate the energies and momenta of the
  left- and right-moving particles. For the short-range interactions, the
  weaker processes of Fig.~\ref{fig:basic-processes}(c) are negligible because
  they do not provide any additional channels of relaxation of the
  system.}\BibitemShut {Stop}%
\end{thebibliography}
\end{document}